\documentclass[%
reprint,
 superscriptaddress,
 amsmath,amssymb,
 aip,
]{revtex4-1}

\usepackage{graphicx}%
\usepackage{dcolumn}%
\usepackage{bm}%
\usepackage{hyperref}%
\usepackage{color}
\usepackage{textcomp}

\newcommand{\eqnref}[1]{Eqn.~(\ref{#1})}
\newcommand{\secref}[1]{Section~\ref{#1}}
\newcommand{\appendixref}[1]{Appendix~\ref{#1}}
\newcommand{\software}[1]{\textsc{#1}}
\newcommand{\term}[1]{#1}

\begin{document}

\preprint{APS/123-QED}

\title{Derivative structure enumeration using binary decision diagram}%

\author{Kohei Shinohara}
    \email{shinohara@cms.mtl.kyoto-u.ac.jp}
    \affiliation{Department of Materials Science and Engineering, Kyoto University, Kyoto 606-8501, Japan}
\author{Atsuto Seko}
    \email{seko@cms.mtl.kyoto-u.ac.jp}
    \affiliation{Department of Materials Science and Engineering, Kyoto University, Kyoto 606-8501, Japan}
    \affiliation{Center for Elements Strategy Initiative for Structure Materials (ESISM), Kyoto University, Kyoto 606-8501, Japan}

\author{Takashi Horiyama}
    \affiliation{Faculty of Information Science and Technology, Sapporo University, Hokkaido 060-0814, Japan}
\author{Masakazu Ishihata}
    \affiliation{NTT Communication Science Laboratories, Keihanna 619‐237, Japan}
\author{Junya Honda}
    \affiliation{Department of Complexity Science and Engineering, Graduate School of Frontier Sciences, The University of Tokyo, Kashiwa 277-8561, Japan}
    \affiliation{RIKEN, Wako 351-0198, Japan}
\author{Isao Tanaka}
    \affiliation{Department of Materials Science and Engineering, Kyoto University, Kyoto 606-8501, Japan}
    \affiliation{Center for Elements Strategy Initiative for Structure Materials (ESISM), Kyoto University, Kyoto 606-8501, Japan}
    \affiliation{Nanostructures Research Laboratory, Japan Fine Ceramics Center, Nagoya 456-8587, Japan}

\date{\today}

\begin{abstract}
A derivative structure is a nonequivalent substitutional atomic configuration derived from a given primitive cell.
The enumeration of derivative structures plays an essential role in searching for the ground states in multicomponent systems.
However, it is computationally difficult to enumerate derivative structures if the number of derivative structures of a target system becomes huge.
In this study, we introduce a novel compact data structure of the zero-suppressed binary decision diagram (ZDD) for enumerating derivative structures much more efficiently.
We show its simple applications to the enumeration of structures derived from the face-centered cubic and hexagonal close-packed lattices in binary, ternary, and quaternary systems.
The present ZDD-based procedure should significantly contribute not only to various computational approaches based on derivative structures but also to a wide range of combinatorial issues in physics and materials science.
\end{abstract}


%

\maketitle

\section{\label{sec:intro}Introduction}

Structure enumeration has played an essential role in performing crystal structure prediction and in understanding crystal structures.
In general, structure enumeration requires a given policy that restricts an entire set of structures in a continuous configuration space to a discrete set of structures.
Such a policy is the atomic substitution of a given structure, and nonequivalent substitutional structures are called ``derivative structures" \cite{Buerger1947}.
Although an efficient algorithm specially developed to find the ground states in multicomponent systems \cite{PhysRevB.94.134424} has recently been found to be applicable when the cluster expansion method \cite{CE1,CE2,CE3} can describe the configurational energy accurately, a set of derivative structures has been commonly used to search for the ground states in multicomponent systems (e.g., Refs.~\onlinecite{PhysRevLett.87.275508,PhysRevB.77.144104,PhysRevB.86.245202}).
A set of derivative structures itself is also of interest from the viewpoint of crystal chemistry because many existing crystal structures of not only intermetallic alloys but also ionic compounds have been interpreted as derivative structures \cite{wells2012structural,muller1993inorganic}.

The well-known P\'olya counting theorem \cite{polya1937} has a long history of being used to count the number of nonequivalent molecule structures \cite{Polya:1987:CEG:26181} and the number of derivative structures \cite{MCLARNAN1981133}, because they can be regarded as graph coloring problems of assigning colors to graph vertices under a given set of permutations.
Recently, Hart and Forcade proposed an efficient procedure to enumerate the derivative structures themselves, not only their total number \cite{Hart2008,Hart2009}.
Their procedure is based on the enumeration of nonequivalent lattices represented by the Hermite normal form (HNF) and the enumeration of labelings using a finitely generated Abelian group given by the Smith normal form (SNF).
Moreover, faster algorithms have also been reported recently \cite{Mustapha2013,Morgan2017}.
They accelerate the enumeration of ternary and quaternary derivative structures by effectively using binary derivative structures that are enumerated in advance.
Therefore, they cannot be efficient for enumerating binary derivative structures.

These procedures are practically sufficient to enumerate derivative structures required to determine the ground-state structures in a binary alloy with a simple lattice such as a face-centered cubic (fcc) one, because each of many intermetallic compounds has a primitive cell composed of up to 24 atoms \cite{steurer2016intermetallics}.
On the other hand, the possible size of periodicity in derivative structures is very restrictive in a binary system with a small number of symmetry operations and a system with three or more components.
In this study, we propose a much more efficient method of enumerating derivative structures.
We employ a compact data structure developed in the algorithm theory for representing a set of combinations.
In particular, we use the zero-suppressed binary decision diagram (ZDD) \cite{Minato1993}, which has been used to enumerate constraint subgraphs of a given graph, such as all paths between two given vertices ($s$-$t$ paths) \cite{knuth2009art,KAWAHARA2017} and spanning trees \cite{inoue2016graphillion}.
For example, ZDD succeeded in enumerating $s$-$t$ paths for the $27 \times 27$ grid graph, the total number of which reaches as many as approximately $10^{163}$ \cite{iwashita2013efficient, oeisA007764}.
The use of ZDD should enable us to significantly increase the possible size of periodicity in derivative structures.

This paper is organized as follows.
Section \ref{sec:terminology} introduces the terminology for representing a derivative structure mathematically.
Section \ref{sec:unique} shows fundamental ideas to eliminate equivalent structures among all possible substitutional ones, following the works of Hart and Forcade \cite{Hart2008,Hart2009}.
Section \ref{sec:ZDD} introduces ZDD and demonstrates how to apply ZDD to the derivative structure enumeration.
The present method employs an isomorphism-eliminated ZDD proposed in Ref.~\onlinecite{Horiyama2018}, which can be applied to the enumeration of binary derivative structures in a straightforward manner.
In addition, we propose a generalization of the isomorphism-eliminated ZDD to extend the scope of application including the derivative structure enumeration in multicomponent systems.
Section~\ref{sec:results} shows the application of the present ZDD-based method to the enumeration of binary, ternary, and quaternary derivative structures from the fcc and hexagonal close-packed (hcp) primitive cells.
Finally, Section~\ref{sec:optional} describes how to append optional constraints on the enumeration of derivative structures along with the present ZDD-based method.


%

\section{\label{sec:terminology}Terminology to represent derivative structures}

\begin{figure}[tb]
    \centering
    \includegraphics[width=\linewidth]{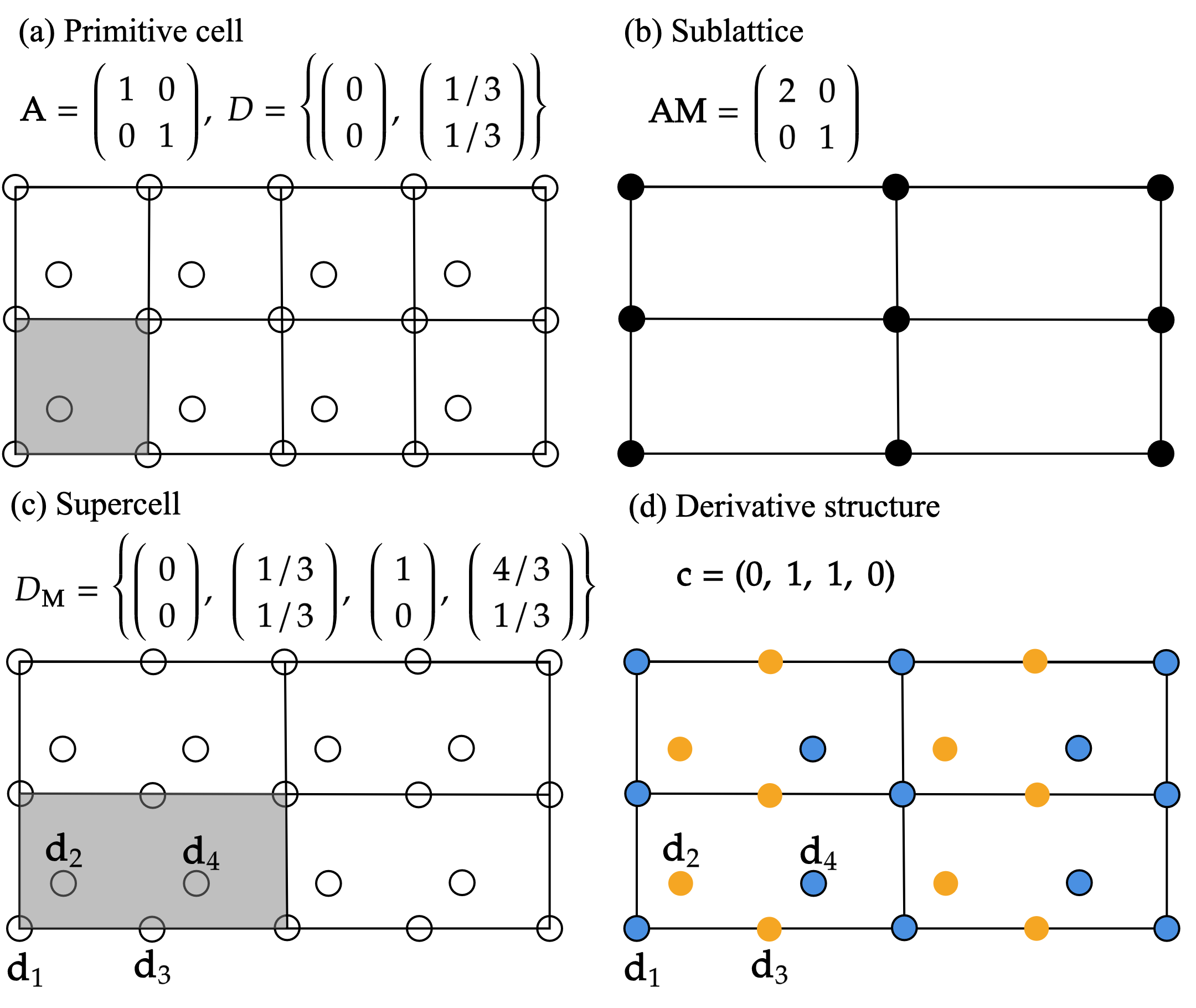}
    \caption{
        (Color online)
        Example of a two-dimensional derivative structure and related terminologies.
        (a)
        A primitive cell with a square parent lattice including two sites and its crystallographic pattern.
        The shaded region represents a primitive cell.
        The point coordinates of the sites in the primitive cell are $D = \{ (0, 0), (1 / 3, 1 / 3) \}$.
        (b)
        A sublattice of the square lattice with the index of two.
        (c)
        A supercell with sublattice (b).
        The shaded region represents a supercell.
        Since the index of sublattice (b) is two, the number of sites in the supercell is twice that in the primitive cell.
        We denote the point coordinates of the sites in the supercell as $D_{\mathbf{M}} = \{\mathbf{d}_{1}, \mathbf{d}_{2}, \mathbf{d}_{3},\mathbf{d}_{4}\}$.
        We use matrix $\mathbf{A}$ as the basis vectors to describe the point coordinates of the sites in the supercell $D_{\mathbf{M}}$.
        (d)
        A binary derivative structure with supercell (c).
        In our label notation, integers 0 and 1 represent the blue and yellow atoms, respectively.
        Therefore, labeling $\mathbf{c} = (0, 1, 1, 0) $ indicates that the blue atoms occupy sites 1 and 4 and the yellow atoms occupy sites 2 and 3.
    }
    \label{fig:terminology}
\end{figure}

In this section, we define the terminology required to represent a derivative structure mathematically.
We mainly follow the convention of crystallography to define them \cite{ITA2016}.
Figure~\ref{fig:terminology} illustrates the terminology in a two-dimensional example, and their mathematical definitions will be given for three dimensions in the following.

\subsection{\label{sec:terminology-unitcell}Parent lattice and primitive cell}

Given a set of basis vectors $\mathbf{A} = \left( \mathbf{a}_{1}, \mathbf{a}_{2}, \mathbf{a}_{3} \right)$, the parent lattice $L$ is defined as
\begin{equation}
L = \left\{ \mathbf{A} \boldsymbol\ell \mid \boldsymbol\ell \in \mathbb{Z}^{3} \right\},
\end{equation}
where $\mathbb{Z}$ denotes the set of integers.
A \term{primitive cell} is also defined as the pair of the set of \term{basis vectors} $\mathbf{A}$ and the set of sites whose positions are described by point coordinates $D$, as shown in Fig.~\ref{fig:terminology}~(a).
We adopt the convention where $\mathbf{A}$ is used as the basis vectors to describe the point coordinates.
Therefore, the values of the point coordinates in the primitive cell range from zero to one.
All of the point coordinates in its crystallographic pattern are expressed as $\{\mathbf{d} + \boldsymbol\ell \mid  \mathbf{d} \in D,\, \boldsymbol\ell \in \mathbb{Z}^{3}\}$.

\subsection{\label{sec:terminology-supercell}Sublattice and supercell}

A \term{sublattice} is a subset of the parent lattice $L$ obtained by removing some lattice points from the parent lattice $L$ \cite{ITA2016}
\footnote{
Although a lattice obtained by removing some translations from the original lattice is called a \term{superlattice} in alloy physics, the term sublattice is mathematically and crystallographically appropriate.
We follow the latter convention throughout this work.
}.
A set of basis vectors of the sublattice is identified with the \term{transformation matrix} $\mathbf{M}$ such that the original set of basis vectors $\mathbf{A}$ is transformed into a new set of basis vectors $\mathbf{AM}$.
Therefore, the sublattice $L_{\mathbf{M}}$ is the set of lattice points expressed as
\begin{equation}
L_{\mathbf{M}} = \left\{ \mathbf{AM} \boldsymbol\ell \mid \boldsymbol\ell \in \mathbb{Z}^{3} \right\}.
\end{equation}
We refer to the determinant of $\mathbf{M}$, $\det \mathbf{M}$, as the \term{index} of the sublattice $L_{\mathbf{M}}$.
The index is identical to the number of lattice points in the sublattice $L_{\mathbf{M}}$
\footnote{
    Throughout this paper, we consider a transformation matrix whose determinant is positive.
}.
Also, a \term{supercell} is identified with the set of basis vectors $\mathbf{AM}$ and the set of point coordinates inside the parallelepiped spanned by the set of basis vectors $\mathbf{AM}$, $D_{\mathbf{M}} = \{ \mathbf{d}_{1}, \dots, \mathbf{d}_{|D_{\mathbf{M}}|} \}$.
The number of sites included in the supercell or the number of point coordinates in $D_{\mathbf{M}}$, $|D_{\mathbf{M}}|$, is given as $|D_{\mathbf{M}}| = |D| \cdot \det \mathbf{M}$.
Figures~\ref{fig:terminology}~(b) and (c) show a sublattice and the corresponding supercell in the two-dimensional example, respectively.

\subsection{\label{sec:terminology-derivative_structure}Derivative structure and labeling}

A $k$-ary derivative structure is defined as a nonequivalent structure in which
one of $k$ atomic species occupies every site of a supercell identified with $\mathbf{AM}$ and $D_\mathbf{M}$.
Therefore, a \term{derivative structure} can be equivalently regarded as the \term{labeling} of the sites for the supercell.
The labeling can be expressed using $k$ integers $\{0, \dots, k - 1 \}$ as
\begin{equation}
    \label{eq:labeling}
    \mathbf{c} = (c_{|D_{\mathbf{M}}|}, c_{|D_{\mathbf{M}}|-1}, \dots, c_{2}, c_{1}) \in \{0, \dots, k - 1 \}^{|D_{\mathbf{M}}|},
\end{equation}
where $c_{i}$ denotes the label of site $i$.
Each label indicates one of the atomic species.
Here we follow the labeling in descending order used in Ref. \onlinecite{Horiyama2018}.
Figure~\ref{fig:terminology}~(d) illustrates a derivative structure in the two-dimensional example.


%
\section{\label{sec:unique}Equivalent structure elimination}

Since the present ZDD-based method follows two fundamental ideas to eliminate equivalent structures, which were used in the successive works of Hart and Forcade \cite{Hart2008, Hart2009}, we summarize them in this section.
One idea is the equivalent sublattice elimination using HNF (\secref{sec:unique-sublattice}).
The other is the equivalent labeling elimination using the structure of the finitely generated Abelian group of a given sublattice (\secref{sec:unique-labeling}).
In \secref{sec:unique-example}, we demonstrate a two-dimensional example of the equivalent labeling elimination for a given sublattice.

\subsection{\label{sec:unique-sublattice}Equivalent sublattice elimination}

As described in the previous section, a sublattice is obtained by transforming basis vectors using the transformation matrix $\mathbf{M}$.
An infinite number of integer transformation matrices are possible even for a given index, which is closely related to the arbitrariness for choosing basis vectors of a given lattice.
Fortunately, however, the number of nonequivalent sublattices is finite for a given index.
We can enumerate a complete set of nonequivalent sublattices for the index.

Let $\mathbf{U}$ be a three-dimensional square \term{unimodular matrix}, where all elements are integers and $\det \, \mathbf{U} = \pm 1$.
It is well known that matrices $\mathbf{M}$ and $\mathbf{MU}$ are equivalent in terms of lattice transformation \cite{Cohen1993}.
This means that they derive the same sublattice expressed as
\begin{equation}
    L_{\mathbf{M}} = L_{\mathbf{MU}},
\end{equation}
although they give different sets of basis vectors spanning the sublattice.
Their representative can be the canonical form called the \term{Hermite normal form} (HNF) \cite{Cohen1993}.
Any transformation matrix $\mathbf{M}$ can be converted to a unique form of the lower-triangular integer matrix, HNF, by multiplying the unimodular matrix $\mathbf{U'}$ from the right satisfying the relationship
\begin{equation}
    \label{eq:HNF}
    \mathbf{MU'} =
    \left(
    \begin{array}{ccc}
         a & 0 & 0 \\
         b & c & 0 \\
         d & e & f
    \end{array}
    \right),
\end{equation}
where $a > 0$, $0 \leq b < c$, $0 \leq d < f$, and $0 \leq e < f$.
The requirement that diagonal elements $a$, $c$, and $f$ are all positive eliminates equivalent basis vectors obtained by inversion.
Also, the addition of a basis vector to another one or the subtraction of a basis vector from another one does not change the lattice itself.
Thus, we can choose remainders of $f$ as $d$ and $e$, and a remainder of $c$ as $b$.

To enumerate nonequivalent sublattices for a given index, therefore, it is sufficient only to enumerate HNFs whose determinant is the index.
HNFs are easily enumerated by brute force.
The product of diagonal elements $a$, $c$, and $f$ should be equal to the index and the diagonal elements should be divisors of the index.
For each set of diagonal elements $\{a, c, f\}$, all combinations of non-diagonal elements $b$, $d$, and $e$ satisfying the inequalities can be generated.
In what follows, we consider only transformation matrices in the lower-triangle HNF.

Then, we eliminate equivalent sublattices among the enumerated ones according to the symmetry of the crystallographic pattern of the primitive cell, as performed in the works of Hart and Forcade~\cite{Hart2008,Hart2009}.
We denote a symmetry operation in the space group of the primitive cell \footnote{
    For simplicity, we abbreviate the space group of the crystallographic pattern generated from a cell as the space group of a cell.
    The crystallographic pattern generated from the cell is different from the cell because the crystallographic pattern is a set of replicas of the cell generated by all translations corresponding to its basis vectors.
    Therefore, the former is more precise than the latter.
}
by the Seitz notation $\{\mathbf{R}| \mathbf{\tau} \}$, where $\mathbf{R}$ and $\mathbf{\tau}$ are a matrix of the point group operation and a vector of the translational operation of the symmetry operation, respectively.
When we choose $\mathbf{A}$ as basis vectors for the matrix $\mathbf{R}$, the point group operation changes the basis vectors of the primitive cell from $\mathbf{A}$ to $\mathbf{AR}$.
Similarly, the point group operation transforms the basis vectors of sublattice $L_{\mathbf{M}}$ from $\mathbf{AM}$ to $\mathbf{ARM}$.
Therefore, if there exists a symmetry operation such that $L_{\mathbf{RM}}$ coincides with a sublattice $L_{\mathbf{M}'}$ that is not $L_{\mathbf{M}}$, two sublattices $L_{\mathbf{M}}$ and $L_{\mathbf{M}'}$ are equivalent.
In other words, $L_{\mathbf{M}}$ and $L_{\mathbf{M}'}$ are equivalent if there exists a unimodular matrix $\mathbf{U}$ such that $\mathbf{RM}$ and $\mathbf{M}'$ satisfy the relationship $\mathbf{RMU} = \mathbf{M}'$; hence, there exists $\mathbf{R}$ such that $(\mathbf{RM})^{-1} \mathbf{M}'$ is unimodular.

\subsection{\label{sec:unique-labeling}Equivalent labeling elimination}
Given the sublattice $L_{\mathbf{M}}$, we enumerate nonequivalent labelings of the sites in the supercell with the sublattice $L_{\mathbf{M}}$.
To eliminate equivalent labelings from all possible labelings, we introduce a permutation representation for symmetry operations of the supercell.
Considering symmetry operations in the space group of the primitive cell that leaves the supercell unchanged, they form a subgroup of the space group of the primitive cell $\mathcal{H}_{\mathbf{M}}$.
If the symmetry operation $g \in \mathcal{H}_{\mathbf{M}}$ moves site $i$ to site $j$ in the supercell, we describe operation $g$ as a permutation, $\sigma_{g}(i) = j$.
By applying this rule to all symmetry operations in $\mathcal{H}_{\mathbf{M}}$, we obtain the permutation group $\Sigma_{\mathbf{M}}$ mapped from the space group $\mathcal{H}_{\mathbf{M}}$ as
\begin{equation}
    \label{eq:permutation_group}
    \Sigma_{\mathbf{M}} = \left\{ \sigma_{g} \mid g \in \mathcal{H}_{\mathbf{M}} \right\},
\end{equation}
which is homomorphic to the space group $\mathcal{H}_{\mathbf{M}}$.

We then define a permutation of labeling $\mathbf{c}$.
Given permutation $\sigma \in \Sigma_{\mathbf{M}}$, permuted labeling $\sigma (\mathbf{c})$ is expressed as
\begin{equation}
    \label{eq:permutation_labeling}
    \sigma (\mathbf{c}) = (c_{ \sigma(|D_{\mathbf{M}}|) }, \dots, c_{ \sigma(1)}).
\end{equation}
We refer to labelings $\mathbf{c}$ and $\sigma (\mathbf{c})$ as equivalent labelings for the supercell with the permutation group $\Sigma_{\mathbf{M}}$.
Therefore, the set of labelings equivalent to $\mathbf{c}$ is given by the orbit of $\mathbf{c}$ as
\begin{equation}
    \label{eq:labeling_orbit}
    \{ \sigma (\mathbf{c}) \mid \sigma \in \Sigma_{\mathbf{M}} \}.
\end{equation}
A representative of the labeling orbit can be defined as the maximum labeling in the lexicographical order, as described in Ref.~\onlinecite{Mustapha2013}.
For example, labeling $\mathbf{c}_{1} = (0, 1, 1, 0)$ is larger in the lexicographical order than labeling $\mathbf{c}_{2} = (0, 0, 1, 1)$.
We denote that $\mathbf{c}_{1}$ is larger than or equal to $\mathbf{c}_{2}$ by $\mathbf{c}_{1} \succeq \mathbf{c}_{2}$.
Finally, nonequivalent labelings for the supercell with the sublattice $L_\mathbf{M}$ can be written as
\begin{equation}
    \label{eq:polya-problem}
    \mathcal{C}_{\mathbf{M}, k} =
    \left\{ \mathbf{c} \in \{0,\dots,k-1\}^{|D_\mathbf{M}|}
        \mid \mathbf{c} \succeq \sigma (\mathbf{c}) \, (\forall \sigma \in \Sigma_{\mathbf{M}})
    \right\}.
\end{equation}
Moreover, \term{P\'olya's counting theorem} is applicable only for determining the size of $\mathcal{C}_{\mathbf{M}, k}$ or counting the number of nonequivalent labelings \cite{polya1937, Polya:1987:CEG:26181} (see \appendixref{sec:appendix-polya}).

The set of nonequivalent labelings depends only on the permutation group.
This means that $\mathcal{C}_{\mathbf{M}, k}$ and $\mathcal{C}_{\mathbf{M'}, k}$ for $\mathbf{M} \neq \mathbf{M'}$ having a bijection or one-to-one correspondence are isomorphic if they have isomorphic permutation groups, i.e., $\Sigma_{\mathbf{M}} \cong \Sigma_{\mathbf{M}'}$.
Therefore, it is sufficient to enumerate nonequivalent labelings only for a complete set of non-isomorphic permutation groups.

Note that the set of nonequivalent labelings given by \eqnref{eq:polya-problem} contains \term{superperiodic} labelings that can be expressed by a smaller supercell and labelings with less than $k$ atomic species denoted as incomplete labelings in Ref.~\onlinecite{Hart2008}.

\subsection{\label{sec:unique-example}Two-dimensional example}

\begin{figure*}[tb]
    \centering
    \includegraphics[width=0.9\linewidth]{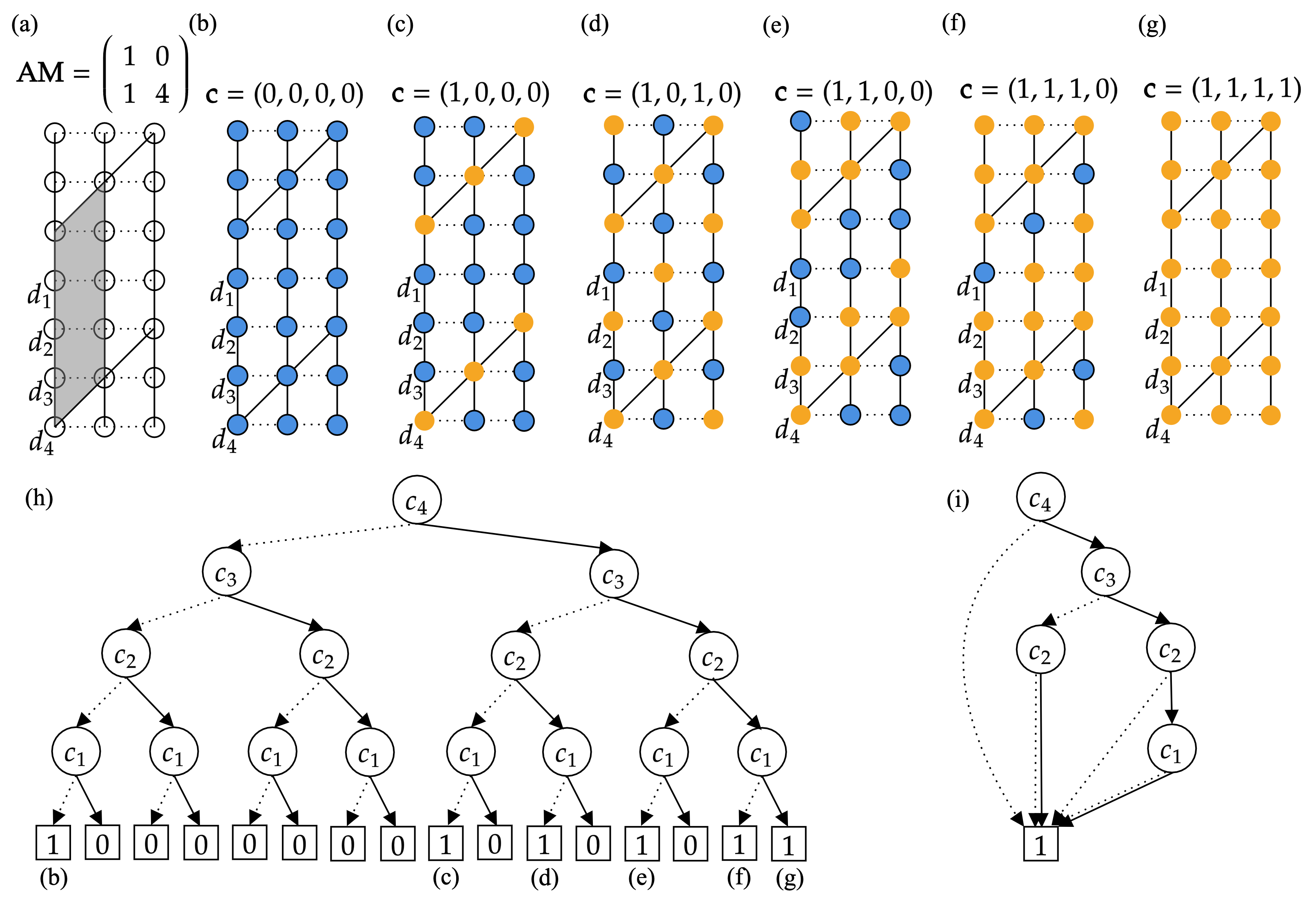}
    \caption{
        (Color online)
        Example of two-dimensional binary derivative structures and their representations.
        (a)
        A sublattice of a square lattice and the corresponding supercell.
        The solid lines indicate the sublattice.
        The point coordinates of the sites in the supercell are denoted by $\mathbf{d}_{1}$, $\mathbf{d}_{2}$, $\mathbf{d}_{3}$, and $\mathbf{d}_{4}$.
        (b)--(g)
        Binary derivative structures and their labelings $\mathbf{c} = (c_4,c_3,c_2,c_1)$.
        The blue contour atoms and yellow atoms correspond to labels 0 and 1, respectively.
        (h)
        Binary decision tree representing nonequivalent labelings (b)--(g).
        The non-terminal node $c_i$ corresponds to label $c_i$ at site $i$ in the supercell.
        The broken and solid arrows from the non-terminal node $c_i$ indicate 0 and 1 assigned to label $c_i$, respectively.
        The square nodes labeled by 1 and 0 indicate that the labeling is the maximum or not the maximum, respectively, among its equivalent labelings in the lexicographic order.
        (i)
        Irreducible ZDD derived from the binary decision tree (h).
    }
    \label{fig:square_lattice}
\end{figure*}

We demonstrate an example of enumerating derivative structures from a two-dimensional primitive cell.
The primitive cell is composed of a square lattice and a site at the origin.
The basis vectors of the primitive cell and the point coordinates of the site are expressed as
\begin{eqnarray}
    \mathbf{A} &= \begin{pmatrix}
        1 & 0 \\
        0 & 1 \\
    \end{pmatrix}, \:\:
    D &= \left\{ \begin{pmatrix} 0 \\ 0 \end{pmatrix} \right\}.
\end{eqnarray}
The space group type of its crystallographic pattern is $p4mm$ \cite{ITA2016}.
We hereafter consider the enumeration of derivative structures for the transformation HNF matrix of
\begin{equation}
    \mathbf{M} = \begin{pmatrix}
        1 & 0 \\
        1 & 4 \\
    \end{pmatrix} .
\end{equation}

Figure~\ref{fig:square_lattice}~(a) illustrates the sublattice $L_{\mathbf{M}}$ and the corresponding supercell.
Since the determinant of the transformation matrix is $\det \mathbf{M} = 4$, the number of sites in the supercell is $|D_{\mathbf{M}}| = 4$.
Although the crystallographic pattern of the primitive cell has a fourfold rotation, that of the supercell does not.
The permutation group of the supercell is represented by the two-line notation \cite{Armstrong} as
\begin{eqnarray}
    \Sigma_{\mathbf{M}} =
        &
        \left\{\begin{pmatrix}
            1 & 2 & 3 & 4 \\
            1 & 2 & 3 & 4
        \end{pmatrix},
        \begin{pmatrix}
            1 & 2 & 3 & 4 \\
            2 & 3 & 4 & 1
        \end{pmatrix},
        \begin{pmatrix}
            1 & 2 & 3 & 4 \\
            3 & 4 & 1 & 2
        \end{pmatrix},
        \right. \nonumber \\&
        \left.
        \begin{pmatrix}
            1 & 2 & 3 & 4 \\
            4 & 1 & 2 & 3
        \end{pmatrix},
        \begin{pmatrix}
            1 & 2 & 3 & 4 \\
            1 & 4 & 3 & 2
        \end{pmatrix},
        \begin{pmatrix}
            1 & 2 & 3 & 4 \\
            3 & 2 & 1 & 4
        \end{pmatrix},
        \right. \nonumber \\&
        \left. \begin{pmatrix}
            1 & 2 & 3 & 4 \\
            2 & 1 & 4 & 3
        \end{pmatrix},
        \begin{pmatrix}
            1 & 2 & 3 & 4 \\
            4 & 3 & 2 & 1
        \end{pmatrix}
    \right\},
\end{eqnarray}
where the first and second rows represent the original sequence and the permuted sequence, respectively.

As described in \appendixref{sec:appendix-polya}, the number of nonequivalent labelings with $k$ atomic species is easily obtained from P\'{o}lya's counting theorem as
\begin{equation}
    \label{eq:example-polya-counting}
    | \mathcal{C}_{\mathbf{M}, k}|
    = \frac{1}{8} \left( k^{4} + 2 k^{3} + 3 k^{2} + 2k \right).
\end{equation}
In the binary case ($k=2$), P\'{o}lya's counting theorem indicates that there are six nonequivalent labelings for the present supercell.
Although the set of all possible labelings is given by $\{\mathbf{c} \mid \mathbf{c} \in \{0, 1\}^{4}\}$, the six nonequivalent labelings are the maximum labelings in the lexicographical order, expressed as
\begin{eqnarray}
    \label{eq:example-labelings-2d}
    \mathcal{C}_{\mathbf{M}, 2} =
    &\left\{
    (0, 0, 0, 0),
    (1, 0, 0, 0),
    (1, 0, 1, 0),
    \right. \nonumber \\
    &\left.
    (1, 1, 0, 0),
    (1, 1, 1, 0),
    (1, 1, 1, 1)
    \right\}.
\end{eqnarray}
They are identical to the nonequivalent labelings shown in Figs.~\ref{fig:square_lattice}~(b)--(g).


%
\section{\label{sec:ZDD}Decision diagram}

In this section, we demonstrate a ZDD-based method to enumerate nonequivalent labelings much more efficiently.
In \secref{sec:ZDD-combination_tree}, we describe the binary decision tree representing a family of subsets from a finite number of elements.
Then we introduce a compact form of the binary decision tree, ZDD, in \secref{sec:ZDD-definition}.
Finally, we propose a procedure to construct a ZDD representing a set of binary nonequivalent labelings (\secref{sec:ZDD-binary}) and a procedure to construct a ZDD representing a set of multicomponent nonequivalent labelings (\secref{sec:ZDD-multi}).

\subsection{\label{sec:ZDD-combination_tree}Binary decision tree}

\begin{figure*}[tb]
    \centering
    \includegraphics[width=\linewidth]{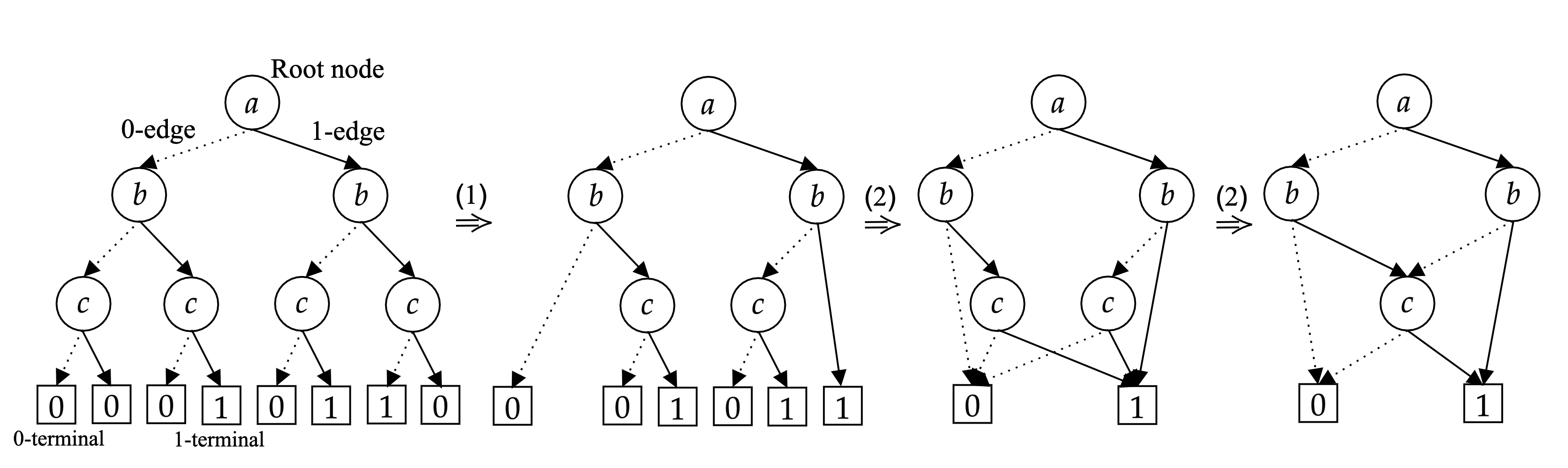}
    \caption{
        Binary decision tree and a process to derive its ZDD for a family of subsets from elements $\{a,b,c\}$.
        The solid and broken arrows indicate 1-edges and 0-edges, respectively.
        The square terminal nodes 1 and 0 indicate 1-terminal and 0-terminal nodes, respectively.
        A reduction process of the binary decision tree to a ZDD is also shown.
        Processes (1) and (2) denote the node elimination rule and the node sharing rule, respectively.
    }
    \label{fig:combination_zdd}
\end{figure*}

A binary decision tree \cite{knuth1978art,CLRS,gross2005graph} represents a family of subsets generated from $n$ elements satisfying given conditions.
For example, subsets containing exactly two elements from among $\{ a, b, c\}$ are $\{a, b\}$, $\{a, c\}$, and $\{b, c\}$.
By fixing the order of the elements, we express the family of the subsets as a binary decision tree.
The first panel of Fig.~\ref{fig:combination_zdd} shows the binary decision tree representing the family of the subsets, $S = \left\{\{a, b\}, \{a, c\}, \{b, c\} \right\}$.
The binary decision tree comprises terminal nodes, non-terminal nodes, and directed edges.
Each non-terminal node has two kinds of outgoing edges, the \term{1-edge} and the \term{0-edge}.
They respectively indicate whether or not a subset includes the element corresponding to the non-terminal node.
Therefore, a path from the root node to a terminal node represents a subset.
Then, the binary value of the terminal node called the \term{1-terminal} or \term{0-terminal} node respectively indicates whether or not the family of subsets $S$ contains the corresponding subset.
Therefore, each of three paths reaching the 1-terminal nodes, called \term{1-paths}, corresponds to each subset in $S = \left\{\{a, b\}, \{a, c\}, \{b, c\} \right\}$.

A binary decision tree also similarly represents a set of nonequivalent labelings.
Figure~\ref{fig:square_lattice}~(h) shows the binary decision tree of the set of nonequivalent labelings in the two-dimensional example shown in Figs.~\ref{fig:square_lattice}~(b)--(g).
The non-terminal node $c_i$ has the \term{1-edge} and the \term{0-edge}, indicating that the corresponding label is assigned as $c_i = 1$ and $c_i = 0$, respectively.
The six 1-paths are identical to the nonequivalent labelings.

\subsection{\label{sec:ZDD-definition}Zero-suppressed binary decision diagram}

\begin{figure}[tb]
    \includegraphics[width=\linewidth]{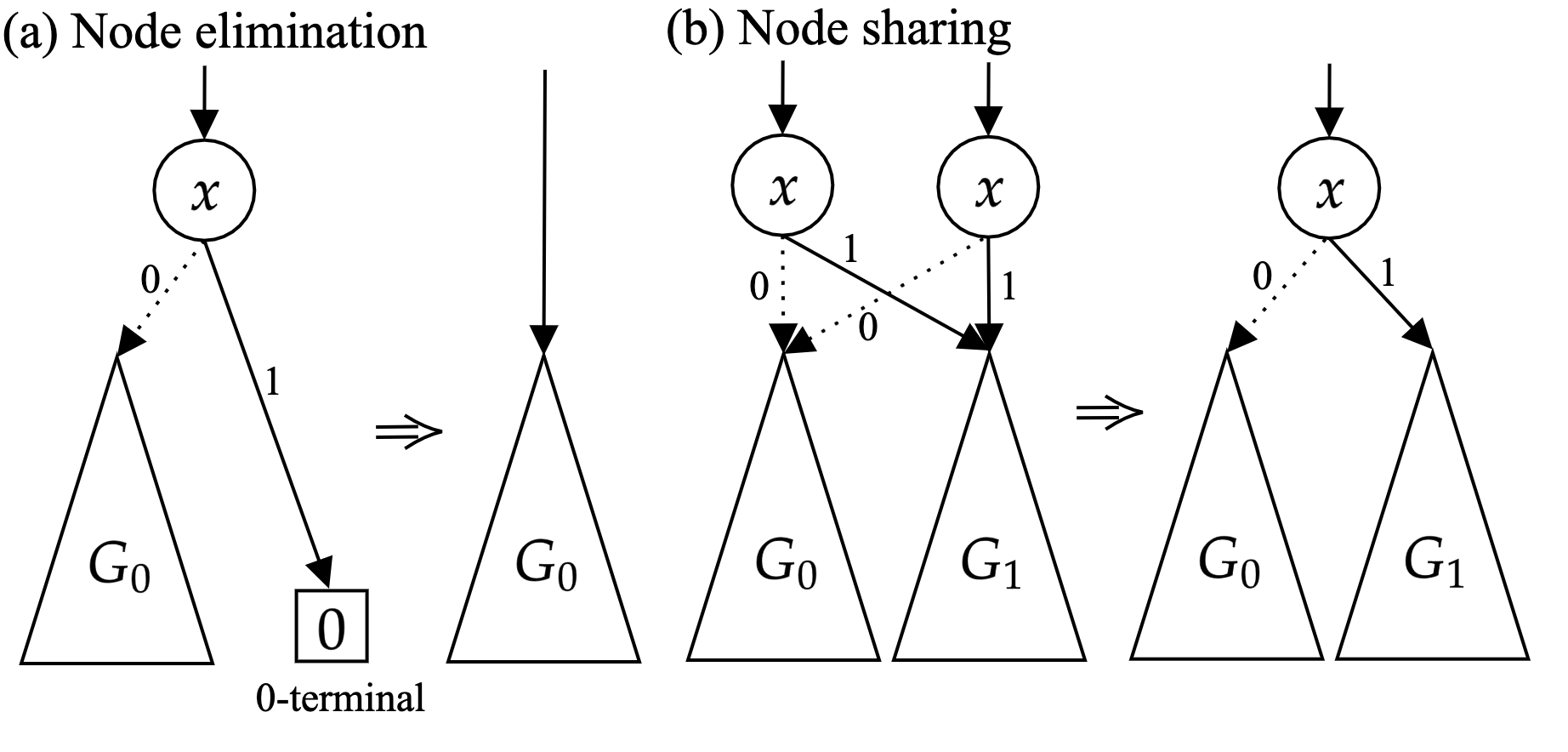}
    \caption{
        ZDD reduction rules.
        (a) Node elimination rule for redundant nodes.
        (b) Node sharing rule for equivalent nodes.
    }
    \label{fig:zdd-rules}
\end{figure}

A binary decision diagram (BDD) is a canonical representation for a Boolean function \cite{bryant1986graph,Bryant1992}, derived by reducing a binary decision tree to a directed acyclic graph.
A ZDD is a variant of the BDD \cite{Minato1993,sasao2014applications} and specially designed for representing sets of combinations.
ZDDs are more efficient than BDDs for representing a family of sparse subsets \cite{Minato1993}.
A ZDD is derived by reducing a binary decision tree on the basis of the following two reduction rules, as schematically illustrated in Fig.~\ref{fig:zdd-rules}.
(1) All nodes whose 1-edge directly points to the 0-terminal nodes are deleted,
and (2) all equivalent nodes having the same child nodes and the same variable are shared.

Figure~\ref{fig:combination_zdd} shows the process of deriving a ZDD from the binary decision tree representing the family of subsets $S = \left\{\{a, b\}, \{a, c\}, \{b, c\} \right\}$.
Fixing the order of variables $a$, $b$, and $c$, two non-terminal nodes $c$ whose 1-edge is directly connected to the 0-terminal nodes are first eliminated, as can be seen in the second panel of Fig.~\ref{fig:combination_zdd}.
Then, redundant terminal nodes are combined into single 1-terminal and 0-terminal nodes, as shown in the third panel of Fig.~\ref{fig:combination_zdd}.
Finally, two non-terminal nodes $c$ with the same child terminal nodes are shared, as shown in the fourth panel of Fig.~\ref{fig:combination_zdd}.
The obtained irreducible ZDD is more compact than the binary decision tree in terms of the number of nodes.
Moreover, the irreducible ZDD is canonical and independent of the order of the reduction processes for a given order of variables.

Each 1-path in a ZDD corresponds to a solution.
The following three 1-paths are identical to the family of subsets $S = \{ \{a, b\}, \{a, c\}, \{b, c\} \}$.
\begin{enumerate}
    \item $a \xrightarrow{1{\rm \mathchar`-edge}}
        b \xrightarrow{1{\rm \mathchar`-edge}} \fbox{1}$
    \item $a \xrightarrow{1{\rm \mathchar`-edge}}
        b \xrightarrow{0{\rm \mathchar`-edge}}
        c \xrightarrow{1{\rm \mathchar`-edge}} \fbox{1}$
    \item $a \xrightarrow{0{\rm \mathchar`-edge}}
        b \xrightarrow{1{\rm \mathchar`-edge}}
        c \xrightarrow{1{\rm \mathchar`-edge}} \fbox{1}$
\end{enumerate}
Note that all paths from the root node to the terminal nodes are represented with a common order of variables.

Figure~\ref{fig:square_lattice}~(i) shows the irreducible ZDD for the set of nonequivalent labelings constructed from the binary decision tree shown in Fig.~\ref{fig:square_lattice}~(h).
The following six 1-paths in the ZDD represent the six nonequivalent labelings.
\begin{enumerate}
    \item $c_4 \xrightarrow{0{\rm \mathchar`-edge}}
        \fbox{1}$
    \item $c_4 \xrightarrow{1{\rm \mathchar`-edge}}
        c_3 \xrightarrow{0{\rm \mathchar`-edge}}
        c_2 \xrightarrow{0{\rm \mathchar`-edge}}
        \fbox{1}$
    \item $c_4 \xrightarrow{1{\rm \mathchar`-edge}}
        c_3 \xrightarrow{0{\rm \mathchar`-edge}}
        c_2 \xrightarrow{1{\rm \mathchar`-edge}}
        \fbox{1}$
    \item $c_4 \xrightarrow{1{\rm \mathchar`-edge}}
        c_3 \xrightarrow{1{\rm \mathchar`-edge}}
        c_2 \xrightarrow{0{\rm \mathchar`-edge}}
        \fbox{1}$
    \item $c_4 \xrightarrow{1{\rm \mathchar`-edge}}
        c_3 \xrightarrow{1{\rm \mathchar`-edge}}
        c_2 \xrightarrow{1{\rm \mathchar`-edge}}
        c_1 \xrightarrow{0{\rm \mathchar`-edge}}
        \fbox{1}$
    \item $c_4 \xrightarrow{1{\rm \mathchar`-edge}}
        c_3 \xrightarrow{1{\rm \mathchar`-edge}}
        c_2 \xrightarrow{1{\rm \mathchar`-edge}}
        c_1 \xrightarrow{1{\rm \mathchar`-edge}}
        \fbox{1}$
\end{enumerate}
They are identical to the set of nonequivalent labelings given by \eqnref{eq:example-labelings-2d}.
In general, the number of paths is calculated by dynamic programming in a computational time proportional to the number of ZDD nodes.
This is typically a fast way to count the number of paths because the number of nodes of a ZDD is much smaller than that of paths in most cases.

\subsection{\label{sec:ZDD-ds}ZDD for derivative structures}

\subsubsection{\label{sec:ZDD-binary}Binary system}

Here, we reformulate the definition of the set of nonequivalent labelings to derive its ZDD.
As described above, a nonequivalent labeling for a given supercell with the sublattice $L_\mathbf{M}$ is defined as the maximum labeling in the lexicographical order among its equivalent labelings for the permutation group $\Sigma_\mathbf{M}$.
In other words, a nonequivalent labeling is a labeling larger than any of its permuted structures for the permutation group.
For permutation $\sigma \in \Sigma_\mathbf{M}$, the set of larger or equal labelings in the lexicographical order is expressed as
\begin{equation}
    \label{eq:permutation-winner}
    \mathcal{C}_{\mathbf{M},2}^{(\sigma)} =
    \left\{ \mathbf{c} \in \{0,1\}^{|D_\mathbf{M}|}
    \mid \mathbf{c} \succeq \sigma(\mathbf{c}) \right\}.
\end{equation}
Therefore, the set of the nonequivalent labelings $\mathcal{C}_{\mathbf{M}, 2}$ is then given by the intersection of $\mathcal{C}_{\mathbf{M},2}^{(\sigma)}$ for all permutations as
\begin{equation}
   \label{eq:zdd-binary}
    \mathcal{C}_{\mathbf{M}, 2} = \bigcap_{\sigma \in \Sigma_{\mathbf{M}}} \mathcal{C}_{\mathbf{M},2}^{(\sigma)}.
\end{equation}
The intersection of ZDDs is efficiently obtained in a top-down manner \cite{Iwashita13} during the construction of an isomorphism-eliminated ZDD, $\mathcal{C}_{\mathbf{M},2}^{(\sigma)}$, using a frontier-based method \cite{Horiyama2018}, as described in Appendix~\ref{sec:frontier}.

\subsubsection{\label{sec:ZDD-multi}Multicomponent system}

We extend the procedure for deriving ZDDs to multicomponent systems, keeping the binary structure of ZDD.
First, we introduce an encoding of labeling $\mathbf{c} \in \{0, \dots, k-1 \}^{|D_{\mathbf{M}}|}$ to one-hot representation $\tilde{\mathbf{c}} \in \{ 0, 1 \}^{k|D_{\mathbf{M}}|}$ expressed as
\begin{eqnarray}
\tilde{\mathbf{c}} & = & (
\tilde{c}_{|D_{\mathbf{M}}|,k-1}, \dots, \tilde{c}_{|D_{\mathbf{M}}|,0},
\dots,
\nonumber \\
& &
\tilde{c}_{2,k-1}, \dots, \tilde{c}_{2,0},
\tilde{c}_{1,k-1}, \dots, \tilde{c}_{1,0}),
\end{eqnarray}
where
\begin{equation}
    \tilde{c}_{i, p} = \begin{cases}
        1 & (c_{i} = p) \\
        0 & (\mathrm{otherwise})
    \end{cases}.
\end{equation}
Labeling $\mathbf{c}$ and one-hot encoding $\tilde{\mathbf{c}}$ have a one-to-one correspondence.
The sum of $k$ elements for site $i$ in the one-hot encoding must be one, because any one of $k$ atomic species occupies site $i$.
This means that the one-hot encoding must satisfy the one-of-$k$ constraint of
\begin{equation}
\label{eq:multi-constraint}
    \sum_{p=0}^{k-1} \tilde{c}_{i, p} = 1 \quad \left( i = 1, \dots, |D_{\mathbf{M}}| \right).
\end{equation}

\begin{figure}
    \centering
    \includegraphics[height=0.5\textheight]{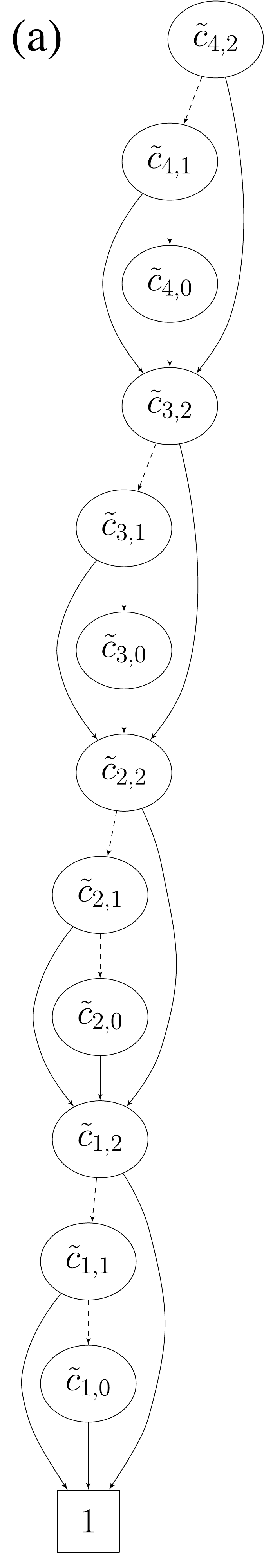}
    \includegraphics[height=0.5\textheight]{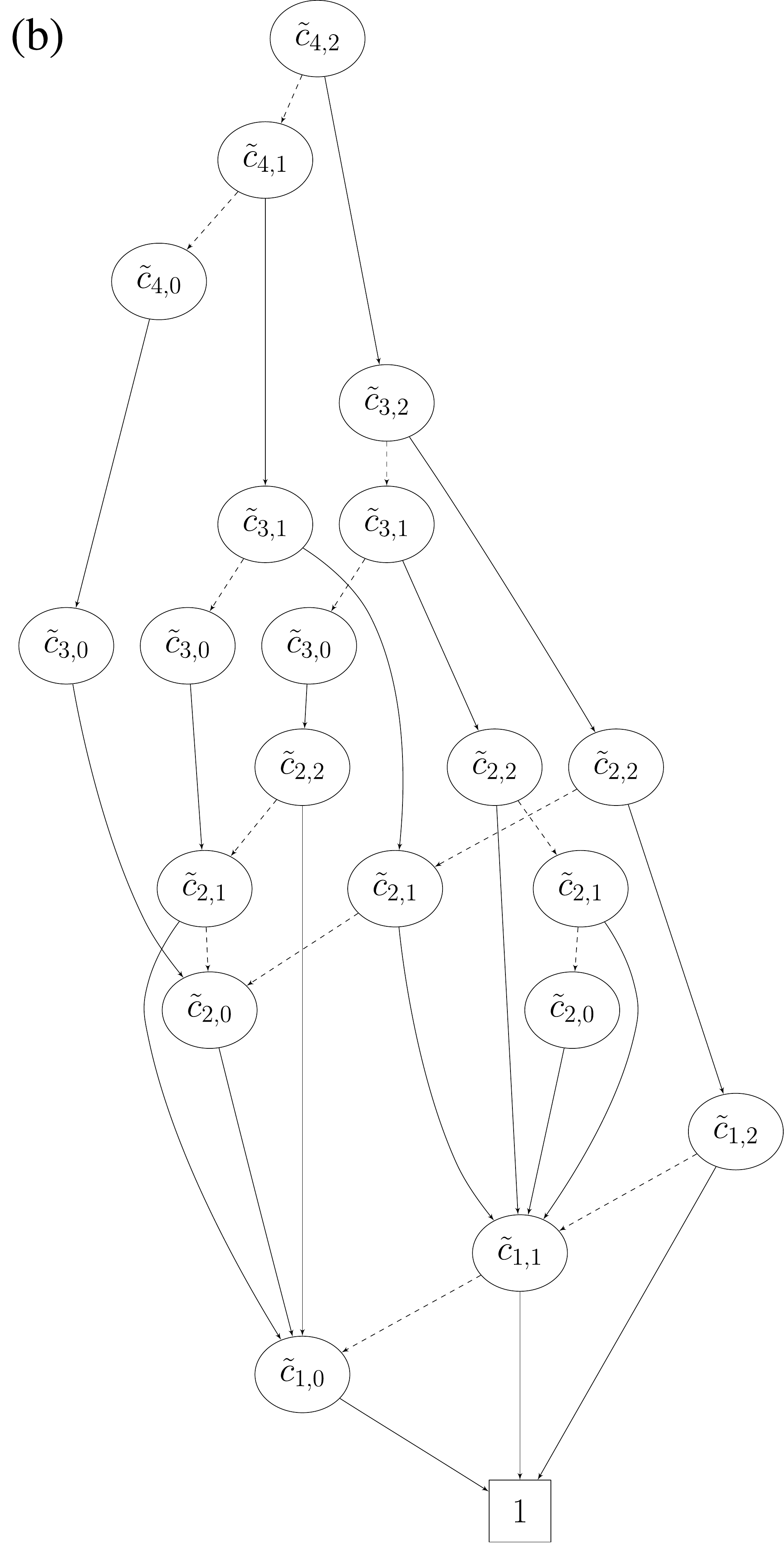}
    \caption{
        (a) One-of-$k$ ZDD of the two-dimensional example representing ternary one-hot encodings that satisfy the one-of-$k$ constraint.
        (b) ZDD representing ternary one-hot encodings of the two-dimensional supercell, which is the same as that used in \secref{sec:unique-example}.
        The solid and broken arrows indicate 1-edges and 0-edges, respectively.
        The square terminal nodes indicate the 1-terminal nodes.
        The 0-terminal node and edges connected to it are omitted for visibility.
    }
    \label{fig:square_k-3_ZDD}
\end{figure}

We then reformulate nonequivalent labelings using the one-hot encoding.
A nonequivalent one-hot encoding can be defined as the largest one-hot encoding among its equivalent one-hot encodings in the lexicographical order.
Therefore, the set of nonequivalent one-hot encodings $\tilde{\mathcal{C}}_{\mathbf{M}, k}$ is written as
\begin{eqnarray}
    \tilde{\mathcal{C}}_{\mathbf{M}, k} & = &
     \biggl\{ \tilde{\mathbf{c}} \in \{0,1\}^{k|D_\mathbf{M}|}
     \mid \tilde{\mathbf{c}} \succeq \sigma(\tilde{\mathbf{c}}) \:\:
     (\forall \sigma \in \Sigma_{\mathbf{M}}), \nonumber \\
     & & \sum_{p=0}^{k-1} \tilde{c}_{i, p} = 1 \:\:(\forall i) \biggr\},
\end{eqnarray}
where the action of permutation $\sigma$ on one-hot encoding $\tilde{\mathbf{c}}$, $\sigma (\tilde{\mathbf{c}})$, is expressed as
\begin{eqnarray}
\sigma(\tilde{\mathbf{c}}) & = & (
\tilde{c}_{\sigma(|D_{\mathbf{M}}|),k-1}, \dots, \tilde{c}_{\sigma(|D_{\mathbf{M}}|),0},
\nonumber \\
& & \dots,
\nonumber \\
& &
\tilde{c}_{\sigma(2),k-1}, \dots, \tilde{c}_{\sigma(2),0},
\nonumber \\
& &
\tilde{c}_{\sigma(1),k-1}, \dots, \tilde{c}_{\sigma(1),0}).
\end{eqnarray}
From this definition of the nonequivalent one-hot encodings, a ZDD of nonequivalent one-hot encodings is the intersection of isomorphism-eliminated ZDDs and a one-of-$k$ ZDD representing whether the constraint is satisfied or not.
Isomorphism-eliminated ZDD for permutation $\sigma$, $\tilde{\mathcal{C}}_{\mathbf{M},k}^{(\sigma)}$, and one-of-$k$ ZDD $\tilde{\mathcal{C}}_{{\rm one\mathchar`-of\mathchar`-}k}$ are written as
\begin{equation}
    \tilde{\mathcal{C}}_{\mathbf{M},k}^{(\sigma)}
    = \left\{ \tilde{\mathbf{c}} \in \{0,1\}^{k|D_\mathbf{M}|}
    \mid \tilde{\mathbf{c}} \succeq \sigma(\tilde{\mathbf{c}}) \right\}
\end{equation}
and
\begin{equation}
    \tilde{\mathcal{C}}_{{\rm one \mathchar`- of \mathchar`-} k}
    = \left\{ \tilde{\mathbf{c}} \in \{0,1\}^{k|D_\mathbf{M}|}
    \mid \sum_{p=0}^{k-1} \tilde{c}_{i, p} = 1 \, (\forall i) \right\},
\end{equation}
respectively.
One-of-$k$ ZDD $\tilde{\mathcal{C}}_{{\rm one \mathchar`- of \mathchar`-} k}$ as illustrated in Fig.~\ref{fig:square_k-3_ZDD}~(a) is easily derived.
Finally, the set of nonequivalent one-hot encodings is given as
\begin{equation}
    \tilde{\mathcal{C}}_{\mathbf{M}, k}
    = \tilde{\mathcal{C}}_{{\rm one \mathchar`- of \mathchar`-} k} \cap \left( \bigcap_{\sigma \in \Sigma_{\mathbf{M}}} \tilde{\mathcal{C}}^{(\sigma)}_{\mathbf{M},k} \right).
\end{equation}
Note that if we impose additional constraints indicating prior knowledge on derivative structures such as energetically prohibited structures, a ZDD satisfying the additional constraints is derived by the intersection of the ZDD $\tilde{\mathcal{C}}_{\mathbf{M}, k}$ and the additional constraint ZDDs.

Figure~\ref{fig:square_k-3_ZDD}~(b) shows a ZDD representing ternary nonequivalent one-hot encodings of the two-dimensional supercell that is the same as that used in \secref{sec:unique-example}.
P\'olya's counting theorem or \eqnref{eq:example-polya-counting} indicates that there exists 21 nonequivalent labelings.
Because a 1-path represents a nonequivalent one-hot encoding, the ZDD has 21 1-paths.
The relationship between paths and nonequivalent one-hot encoding can be seen in the following example.
Labeling $\mathbf{c} = (2, 1, 0, 0)$ is encoded to a one-hot representation as
\begin{eqnarray}
    \tilde{\mathbf{c}} &=&
    (\tilde{c}_{4, 2}, \tilde{c}_{4, 1}, \tilde{c}_{4, 0},
    \tilde{c}_{3, 2}, \tilde{c}_{3, 1}, \tilde{c}_{3, 0},
    \tilde{c}_{2, 2}, \tilde{c}_{2, 1}, \tilde{c}_{2, 0},
    \tilde{c}_{1, 2}, \tilde{c}_{1, 1}, \tilde{c}_{1, 0})
    \nonumber \\
    &=&
    (1, 0, 0,
     0, 1, 0,
     0, 0, 1,
     0, 0, 1),
\end{eqnarray}
which is stored in the ZDD as the following 1-path.
\begin{itemize}
  \item $\tilde{c}_{4,2} \xrightarrow{\rm{1}} \tilde{c}_{3,2} \xrightarrow{\rm{0}} \tilde{c}_{3,1}  \xrightarrow{\rm{1}} \tilde{c}_{2,2}
\xrightarrow{\rm{0}} \tilde{c}_{2,1} \xrightarrow{\rm{0}} \tilde{c}_{2,0} \xrightarrow{\rm{1}} \tilde{c}_{1,1} \xrightarrow{\rm{0}} \tilde{c}_{1,0} \xrightarrow{\rm{1}} \fbox{1}$
\end{itemize}
Note that the ZDD shown in Fig.~\ref{fig:square_k-3_ZDD}~(b) is much more compact than the corresponding binary decision tree.
The width of the ZDD or the maximum number of nodes corresponding to the same variable is only three, which corresponds to the number of nodes for $\tilde{c}_{3,0}$, $\tilde{c}_{2,2}$, and $\tilde{c}_{2,1}$.
On the other hand, the width of the binary decision tree is $k^{|D_{\mathbf{M}}|} = 3^{4} = 81$.


%

\section{\label{sec:results}Results and Discussion}

\begin{table*}[tbp]
  \begin{ruledtabular}
  \caption{
  Number of nonequivalent sublattices and permutation groups for fcc.
  The notations $N_{\rm HNF}$ and $N_{\Sigma_\mathbf{M}}$ denote the number of nonequivalent sublattices and the number of non-isomorphic permutation groups for a given index.
  }
  \label{table:fcc_hnf_perms}
  \begin{tabular}{ccc|ccc|ccc|ccc}
    Index &  $N_{\rm HNF}$ &  $N_{\Sigma_\mathbf{M}}$ &
    Index &  $N_{\rm HNF}$ &  $N_{\Sigma_\mathbf{M}}$ &
    Index &  $N_{\rm HNF}$ &  $N_{\Sigma_\mathbf{M}}$ &
    Index &  $N_{\rm HNF}$ &  $N_{\Sigma_\mathbf{M}}$ \\
    \hline
    2 &    2 &   1  & 14 &   28 &   2 &  26 &   72 &   3 &  38 &  136 &   2 \\
    3 &    3 &   1  & 15 &   31 &   3 &  27 &   75 &  12 &  39 &  129 &   4 \\
    4 &    7 &   3  & 16 &   58 &  16 &  28 &  123 &   8 &  40 &  286 &  22 \\
    5 &    5 &   2  & 17 &   21 &   2 &  29 &   49 &   2 &  41 &   89 &   2 \\
    6 &   10 &   1  & 18 &   60 &  10 &  30 &  158 &   3 &  42 &  268 &   3 \\
    7 &    7 &   2  & 19 &   25 &   2 &  31 &   55 &   2 &  43 &   97 &   2 \\
    8 &   20 &   7  & 20 &   77 &   8 &  32 &  177 &  33 &  44 &  249 &   6 \\
    9 &   14 &   5  & 21 &   49 &   3 &  33 &   97 &   2 &  45 &  218 &  19 \\
    10 &   18 &  2  & 22 &   54 &   1 &  34 &  112 &   2 &  46 &  190 &   1 \\
    11 &   11 &  1  & 23 &   33 &   1 &  35 &   99 &   4 &  47 &  113 &   1 \\
    12 &   41 &  6  & 24 &  144 &  16 &  36 &  268 &  33 &  48 &  496 &  53 \\
    13 &   15 &  3  & 25 &   50 &   7 &  37 &   75 &   3 &     &      &     \\
  \end{tabular}
  \end{ruledtabular}
\end{table*}

\begin{table}[htbp]
  \begin{ruledtabular}
  \caption{
  Number of nonequivalent sublattices and permutation groups for hcp.
  }
  \label{table:hcp_hnf_perms}
  \begin{tabular}{ccc|ccc}
    Index &  $N_{\rm HNF}$ &  $N_{\Sigma_\mathbf{M}}$ &
    Index &  $N_{\rm HNF}$ &  $N_{\Sigma_\mathbf{M}}$ \\
    \hline
    1 &     1 &    1  & 14 &  53 &   13 \\
    2 &     3 &    2  & 15 &  55 &   20 \\
    3 &     5 &    3  & 16 & 104 &   57 \\
    4 &    11 &    9  & 17 &  37 &   10 \\
    5 &     7 &    4  & 18 & 115 &   32 \\
    6 &    19 &    9  & 19 &  45 &   12 \\
    7 &    11 &    6  & 20 & 143 &   48 \\
    8 &    34 &   25  & 21 &  91 &   27 \\
    9 &    23 &   12  & 22 & 105 &   16  \\
    10 &   33 &   10  & 23 &  61 &   13 \\
    11 &   19 &    7  & 24 & 272 &  108 \\
    12 &   77 &   40  & 25 &  90 &   22 \\
    13 &   25 &    9  &   & &
  \end{tabular}
  \end{ruledtabular}
\end{table}

We demonstrate applications of the present ZDD-based method to the enumeration of binary, ternary, and quaternary derivative structures from the fcc and hcp primitive cells.
Basis vectors of a primitive cell and point coordinates are given as
\begin{equation}
  \mathbf{A}=
    \begin{pmatrix}
      0 & 1 & 1 \\
      1 & 0 & 1 \\
      1 & 1 & 0
    \end{pmatrix}, \:\:
    D=
    \left\{
      \begin{pmatrix} 0 \\ 0 \\ 0 \end{pmatrix}
    \right\},
\end{equation}
and
\begin{equation}
  \mathbf{A}=
  \begin{pmatrix}
    1 & 1/2 & 0 \\
    0 & \sqrt{3}/2 & 0 \\
    0 & 0 & 2\sqrt{6}/3
  \end{pmatrix},  \:\:
  D=
  \left\{
    \begin{pmatrix} 0 \\ 0 \\ 0 \end{pmatrix},
    \begin{pmatrix} 1/3 \\ 1/3 \\ 1/2 \end{pmatrix}
  \right\}
\end{equation}
for fcc and hcp, respectively.
On the basis of the fcc and hcp primitive cells, we enumerate nonequivalent sublattices identified by HNFs and non-isomorphic permutation groups to derive ZDDs representing nonequivalent labelings.
We used \software{Spglib} \cite{spglib} to obtain symmetry operations in the space groups of the primitive cells.
Tables \ref{table:fcc_hnf_perms} and \ref{table:hcp_hnf_perms} show the numbers of nonequivalent sublattices and non-isomorphic permutation groups for fcc and hcp, respectively.
The sequence of nonequivalent sublattices for fcc is found in the On-line Encyclopedia of Integer Sequences (OEIS) (A159842) \cite{Hart2008,oeis}.
The sequence of nonequivalent sublattices for hcp are also found in OEIS (A300783).

\begin{figure}[tb]
  \includegraphics[clip,height=0.25\textheight]{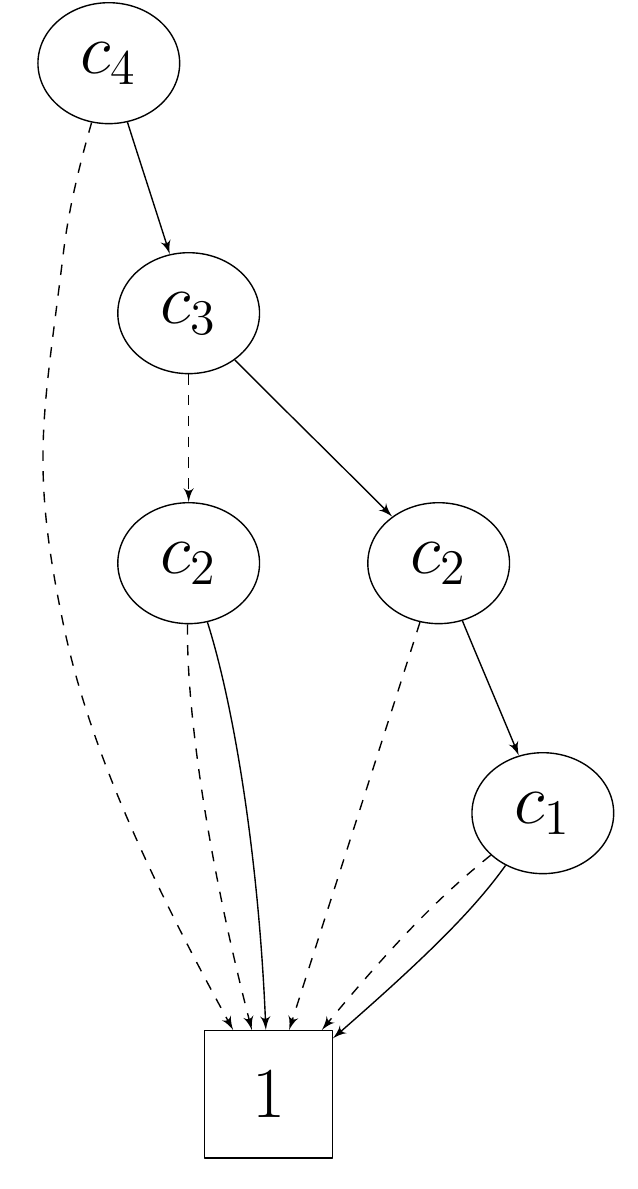}
  \includegraphics[clip,height=0.25\textheight]{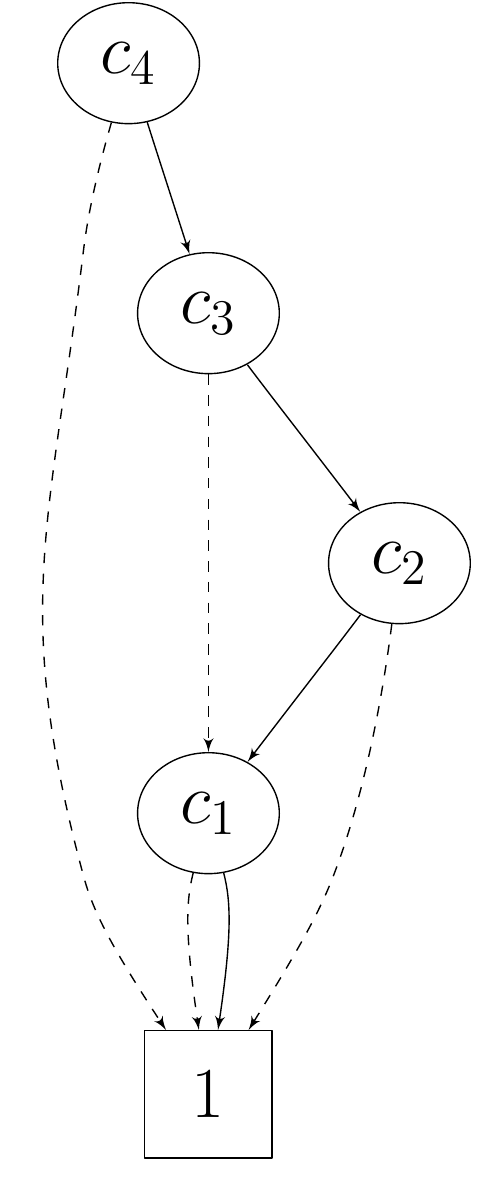}
  \includegraphics[clip,height=0.25\textheight]{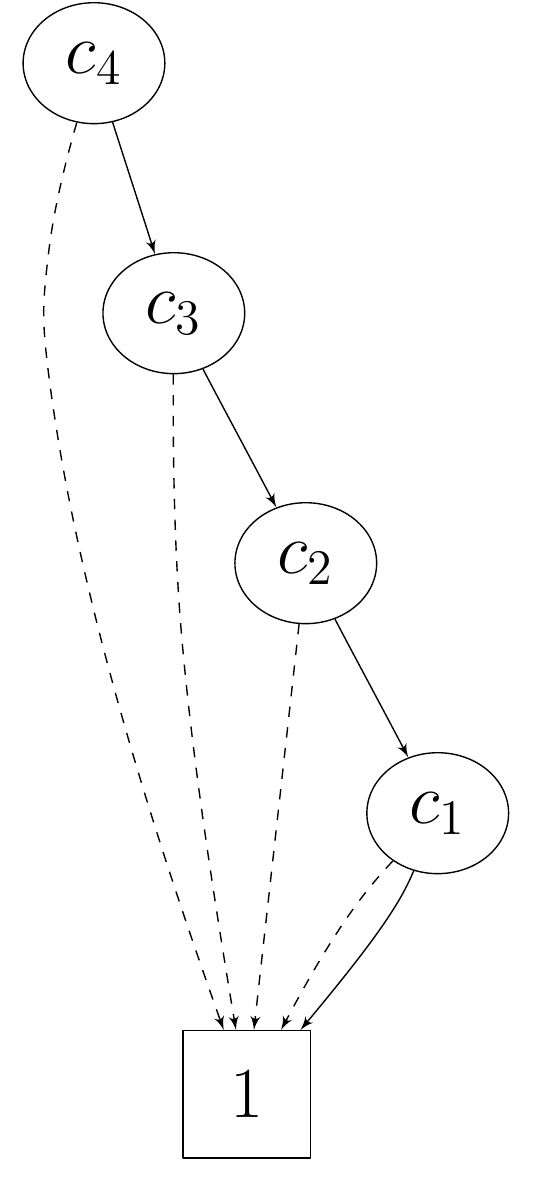}
  \caption{Three ZDDs representing binary fcc nonequivalent labelings with the index of four.
    Each ZDD corresponds to nonequivalent labelings for each non-isomorphic permutation group.
  }
  \label{fig:FCC_ZDDs}
\end{figure}

Then, we derive a ZDD representing nonequivalent labelings for each non-isomorphic permutation group.
We used \software{tdzdd} \cite{TdZdd,Iwashita13}, which is a library for facilitating the manipulation processes of constructing ZDDs by a frontier-based method.
Figure~\ref{fig:FCC_ZDDs} shows ZDDs representing binary fcc nonequivalent labelings with the index of four.
Each ZDD corresponds to nonequivalent labelings for each non-isomorphic permutation group.
The ZDD shown in the left panel of Fig.~\ref{fig:FCC_ZDDs} is identical to a set of nonequivalent labelings for the supercell specified by the following five HNFs:
\begin{eqnarray}
   \mathbf{M} & = &
     \begin{pmatrix} 1 & 0 & 0 \\ 0 & 1 & 0 \\  0 & 0 & 4 \end{pmatrix},
     \begin{pmatrix} 1 & 0 & 0 \\ 0 & 1 & 0 \\ 0 & 1 & 4 \end{pmatrix},
     \begin{pmatrix} 1 & 0 & 0 \\ 0 & 1 & 0 \\ 0 & 2 & 4 \end{pmatrix},
     \nonumber \\
     & &
     \begin{pmatrix} 1 & 0 & 0 \\ 0 & 1 & 0 \\ 0 & 3 & 4 \end{pmatrix},
     \begin{pmatrix} 1 & 0 & 0 \\ 0 & 1 & 0 \\ 1 & 2 & 4 \end{pmatrix}.
\end{eqnarray}
The ZDDs shown in the middle and right panels of Fig.~\ref{fig:FCC_ZDDs} represent nonequivalent labelings for
\begin{equation}
   \mathbf{M} =
    \begin{pmatrix}
      1 & 0 & 0 \\
      0 & 2 & 0 \\
      0 & 0 & 2
    \end{pmatrix}
\end{equation}
and
\begin{equation}
   \mathbf{M} =
    \begin{pmatrix}
      1 & 0 & 0 \\
      1 & 2 & 0 \\
      1 & 0 & 2
    \end{pmatrix},
\end{equation}
respectively.
The last transformation matrix generates the fcc conventional unit cell, and the following two 1-paths correspond to the L1$_2$ (Cu$_3$Au-type) structure.
\begin{itemize}
  \item $c_{4} \xrightarrow{1\mathchar`-{\rm edge}}
         c_{3} \xrightarrow{0\mathchar`-{\rm edge}}
         \fbox{1}$
  \item $c_{4} \xrightarrow{1\mathchar`-{\rm edge}}
         c_{3} \xrightarrow{1\mathchar`-{\rm edge}}
         c_{2} \xrightarrow{1\mathchar`-{\rm edge}}
         c_{1} \xrightarrow{0\mathchar`-{\rm edge}}
         \fbox{1}$
\end{itemize}

\begin{figure}[tb]
  \includegraphics[clip,height=0.25\textheight]{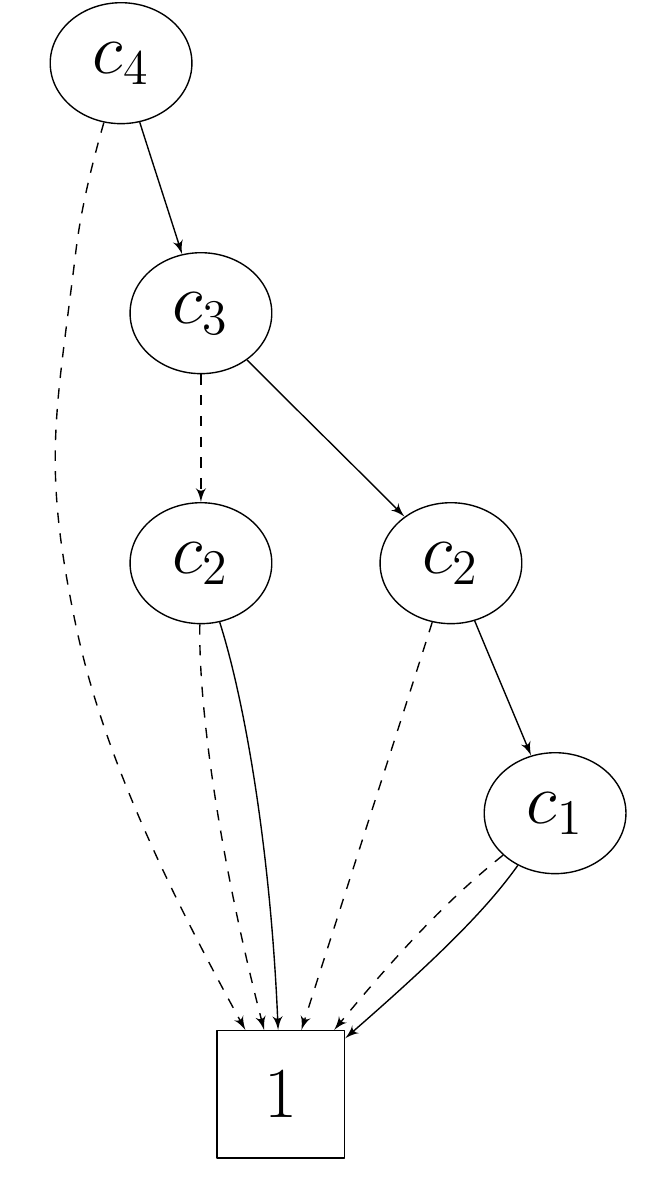}
  \includegraphics[clip,height=0.25\textheight]{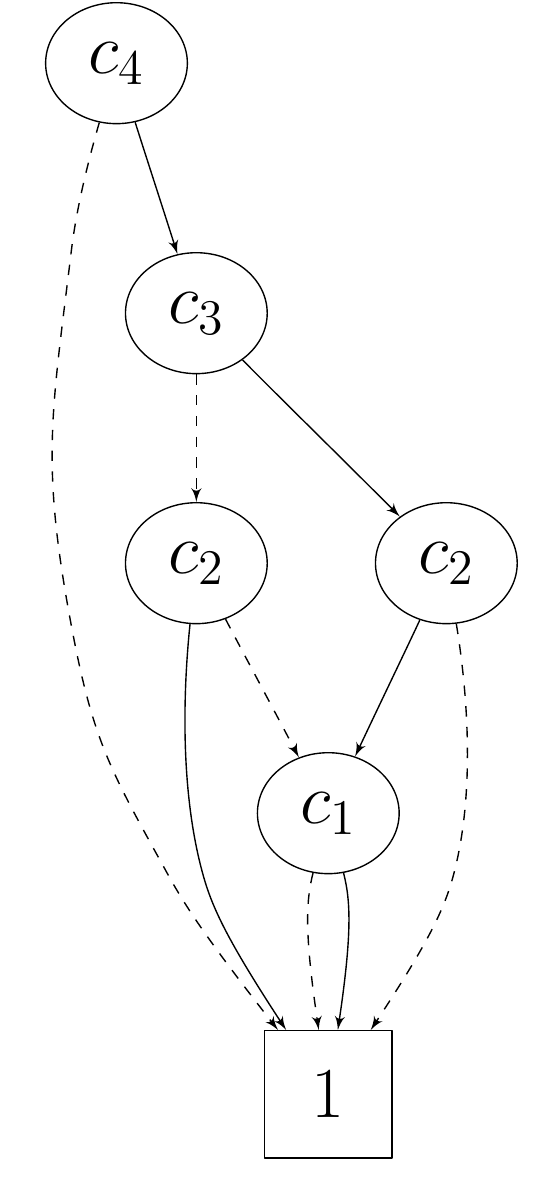}
  \caption{Two ZDDs representing binary hcp nonequivalent labelings with the index of two.}
  \label{fig:HCP_ZDDs}
\end{figure}

Figure~\ref{fig:HCP_ZDDs} shows ZDDs representing binary hcp nonequivalent labelings with the index of two.
The ZDDs shown in the left and right panels of Fig.~\ref{fig:HCP_ZDDs} represent nonequivalent labelings for
\begin{equation}
   \mathbf{M} =
   \begin{pmatrix} 1 & 0 & 0 \\ 0 & 1 & 0 \\  0 & 0 & 2 \end{pmatrix},
   \begin{pmatrix} 1 & 0 & 0 \\ 0 & 1 & 0 \\ 0 & 1 & 2 \end{pmatrix}
\end{equation}
and
\begin{equation}
   \mathbf{M} =
    \begin{pmatrix}
      1 & 0 & 0 \\
      0 & 2 & 0 \\
      0 & 0 & 1
    \end{pmatrix},
\end{equation}
respectively.

\begin{figure*}[tb]
    \centering
    \includegraphics[width=\linewidth]{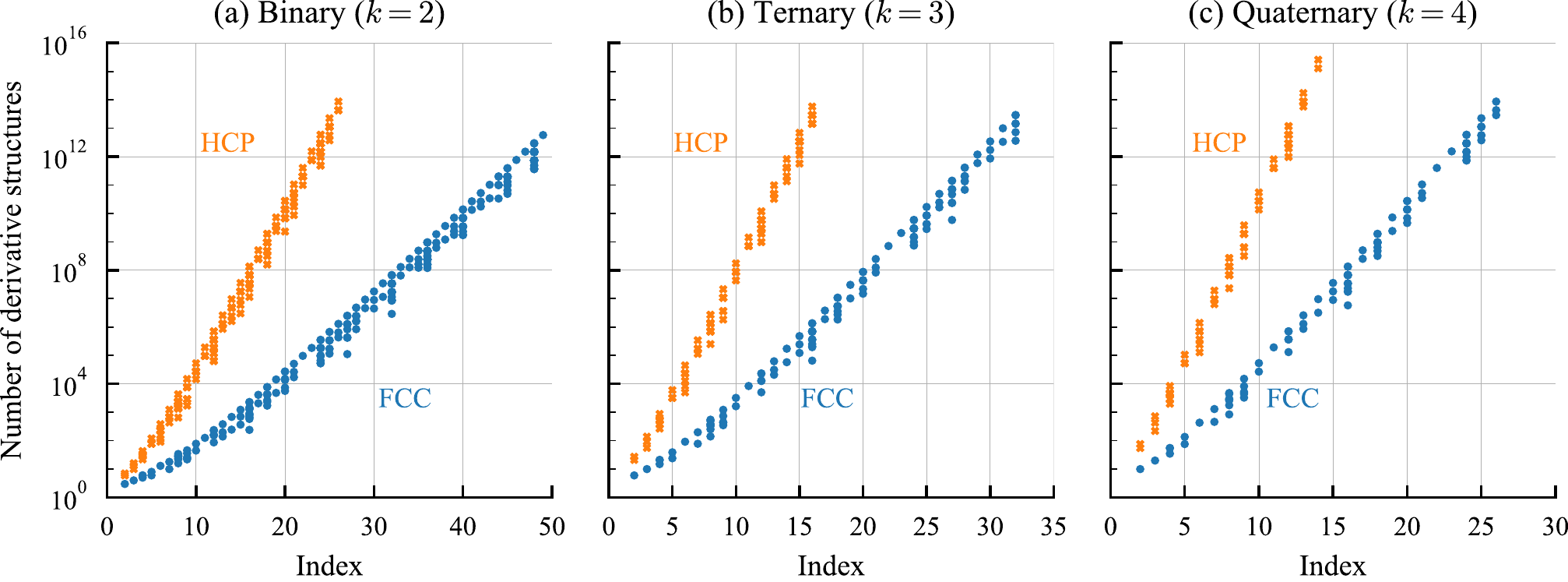}
    \caption{
        (Color online)
        Number of nonequivalent labelings for every permutation group in (a) binary, (b) ternary, and (c) quaternary systems.
        The horizontal axis indicates the index.
        The blue closed circles and orange crosses show the numbers of fcc nonequivalent labelings and hcp nonequivalent labelings, respectively.
    }
    \label{fig:total_counts_fcc_hcp}
\end{figure*}

Then, we compare the performance of the present ZDD-based method with that of the previous method \cite{Hart2008, Hart2009, Morgan2017} implemented in \software{enumlib} \cite{enumlib}.
As performed in the present ZDD-based method, the previous method (\software{enumlib}) enumerates nonequivalent labelings only for non-isomorphic permutation groups for a given index.
We enumerate nonequivalent labelings to the limit of the index due to the computational resource
\footnote{
We used a workstation powered by Intel\textregistered\ Xeon\textregistered\ Processor E5-2695 v4 (2.10~GHz) with 512~GB RAM.
We compiled our C\texttt{++}-based source code using GNU Complier Collection 7.4 with the optimization O3 flag.
We used only a CPU thread for each enumeration.
}.

Figure~\ref{fig:total_counts_fcc_hcp} shows the number of nonequivalent labelings for each non-isomorphic permutation group in (a) binary, (b) ternary, and (c) quaternary systems.
As described before, the number of nonequivalent labelings is easily calculated by P\'olya's counting theorem \cite{polya1937, Polya:1987:CEG:26181}, and the number of nonequivalent labelings obtained from each ZDD coincides exactly with the number obtained by P\'olya's counting theorem.
The multiple numbers of nonequivalent labelings are found at most of the indexes in Fig.~\ref{fig:total_counts_fcc_hcp} because the number of nonequivalent labelings depends on the permutation group.
Note that all nonequivalent labelings themselves can be obtained by tracing all paths in a ZDD, although we show only the number of nonequivalent labelings in Fig.~\ref{fig:total_counts_fcc_hcp}.

For fcc, the present ZDD-based method enumerates nonequivalent labelings with up to 48, 31, and 26 sites for binary, ternary, and quaternary systems, respectively, which are much larger than the possible number of sites using the previous method implemented in \software{enumlib}.
The total number of derivative structures, which is the sum of the numbers of nonequivalent labelings over nonequivalent supercells, reaches approximately $10^{17}$ in each of the binary, ternary, and quaternary systems.
On the other hand, the total number of derivative structures is approximately $10^{10}$ in every system when using the previous method, which is much smaller than that of the present ZDD-based method.
For hcp, the number of sites in a derivative structure is twice its index.
As can be seen in Fig.~\ref{fig:total_counts_fcc_hcp}, the present ZDD-based method enumerates nonequivalent labelings with up to 50, 30, and 26 sites for the binary, ternary, and quaternary systems, respectively.
The total number of derivative structures ranges approximately from $10^{16}$ to $10^{18}$.
Although the number of enumerated derivative structures can be too enormous to use them as candidates for optimization, it is expected to reduce candidates by introducing some useful constraints.

\begin{figure}[tb]
    \centering
    \includegraphics[width=\linewidth]{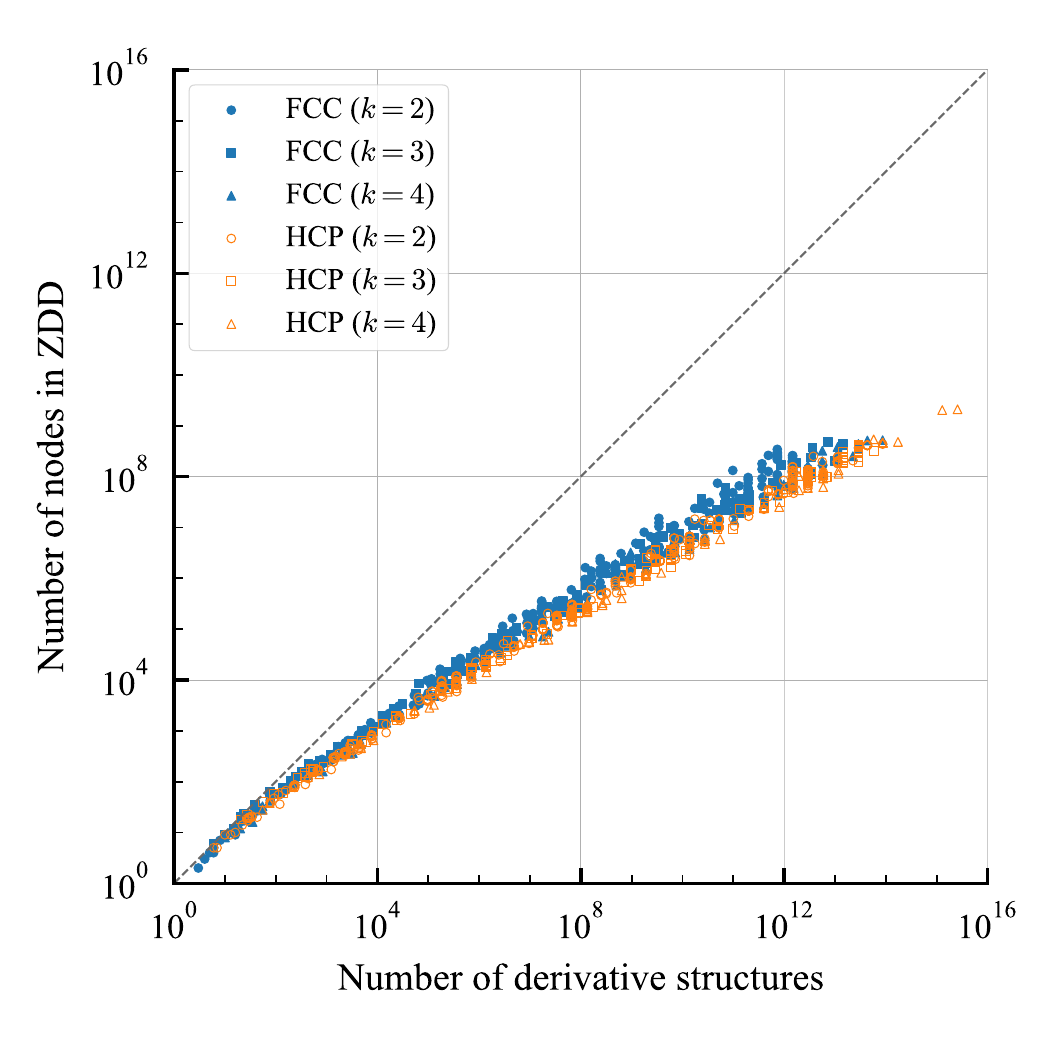}
    \caption{
        (Color online)
        Number of non-terminal nodes in relation to number of nonequivalent labelings in ZDDs.
        The blue and orange symbols indicate ZDDs representing fcc and hcp nonequivalent labelings, respectively.
    }
    \label{fig:nodes_and_confs}
\end{figure}

Figure~\ref{fig:nodes_and_confs} shows the number of non-terminal nodes and the number of nonequivalent labelings in ZDDs.
The diagonal line indicates that the number of non-terminal nodes is equal to that of nonequivalent labelings, which corresponds to a simple case of each labeling being expressed by a single node.
Therefore, their ratio can be a simple estimation of the efficiency of a ZDD.
As can be seen in Fig.~\ref{fig:nodes_and_confs}, the number of non-terminal nodes is much smaller than that of nonequivalent labelings, which indicates that ZDD represents nonequivalent labelings efficiently.
For example, ZDD compresses as many as approximately $10^{12}$ nonequivalent labelings into approximately $10^{8}$ non-terminal nodes.

\begin{figure*}[tb]
    \centering
    \includegraphics[width=\linewidth]{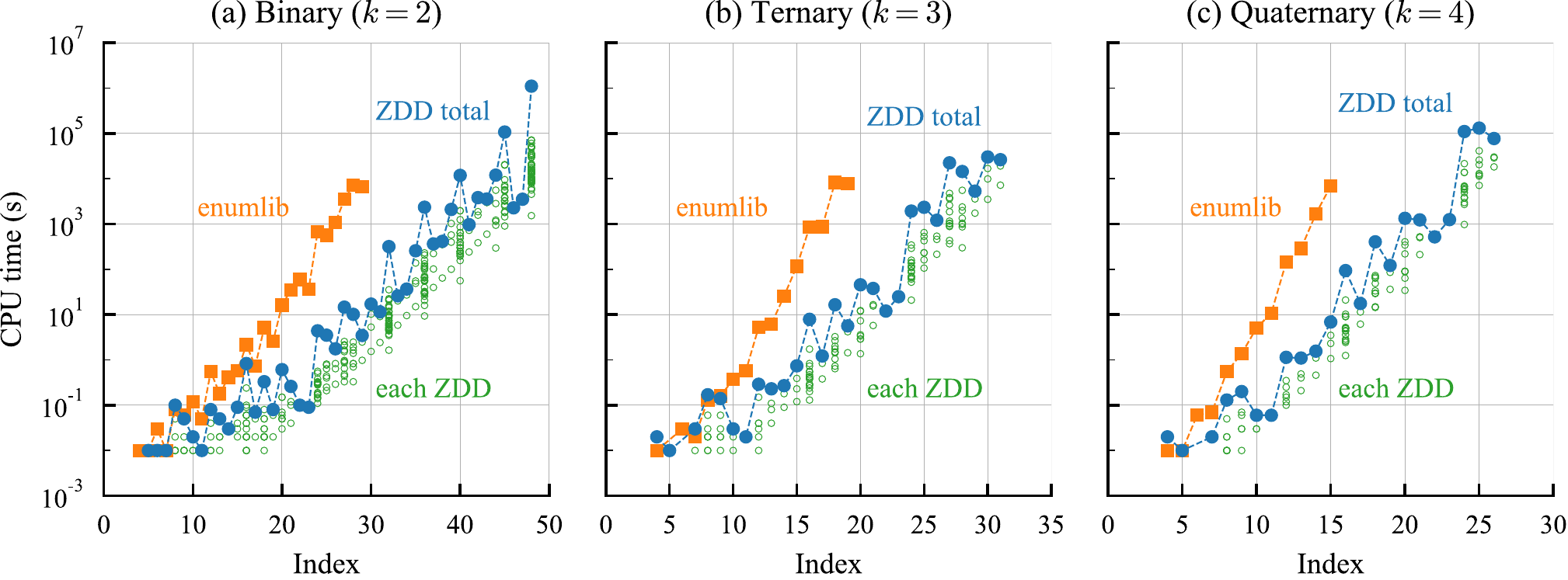}
    \caption{
        (Color online)
        Computational times required to enumerate FCC derivative structures in present ZDD-based and previous methods in (a) binary, (b) ternary, and (c) quaternary systems.
        The blue closed circles stand for constructing ZDDs for each index.
        The green open circles indicate the computational time required to construct a ZDD representing nonequivalent labelings for each permutation group with a given index.
        The computational time of the previous method of enumerating derivative structures for each index is also shown by the orange closed squares.
    }
    \label{fig:zdd_total_vs_enumlib_fcc_time}
\end{figure*}

Figure~\ref{fig:zdd_total_vs_enumlib_fcc_time} shows the computational time required to enumerate derivative structures for a given index using the present ZDD-based method and the previous method (\software{enumlib}).
The computational time required to construct a ZDD for each non-isomorphic permutation group is also shown in Fig.~\ref{fig:zdd_total_vs_enumlib_fcc_time}.
In the ZDD-based method, the computational time required to enumerate derivative structures for a given index is the sum of the computational times required to construct ZDDs for non-isomorphic permutation groups.
Similarly to the present method, the previous method (\software{enumlib}) enumerates nonequivalent labelings only for non-isomorphic permutation groups for a given index.

As can be seen in Fig.~\ref{fig:zdd_total_vs_enumlib_fcc_time}, both methods require exponential time with respect to the index.
However, the two series of computational time indicate that the base of exponential time in the ZDD-based method is half that in the previous method.
For example, the computational time is approximately $10^4$ and $10^1$ seconds at the index of 29 in the previous method and the ZDD-based method, respectively.
The difference in the computational time at a larger index between the previous method and the ZDD-based method is expected to be much larger.
Thus, the ZDD-based method to enumerate derivative structures is much more efficient than the previous method.
In practice, memory consumption is also an essential aspect in enumerating derivative structures.
A comparison of memory consumption between the ZDD-based method and the previous method is given in \appendixref{sec:memory}.

Note that it is impossible to compare the present ZDD-based method and the previous method in a rigorous manner, because they have some differences in their procedures and implementations.
However, the present discussion on their differences in terms of computational time order should remain valid.
The main difference between their procedures is that the previous method (\software{enumlib}) excludes superperiodic structures and incomplete structures, but the present method includes such structures.
Regarding the implementation of the methods, the present method and the previous method (\software{enumlib}) are implemented in C\texttt{++} and Fortran, respectively.


%
\section{\label{sec:optional}Optional constraints}

\begin{figure*}[tb]
    \centering
    \includegraphics[width=\linewidth]{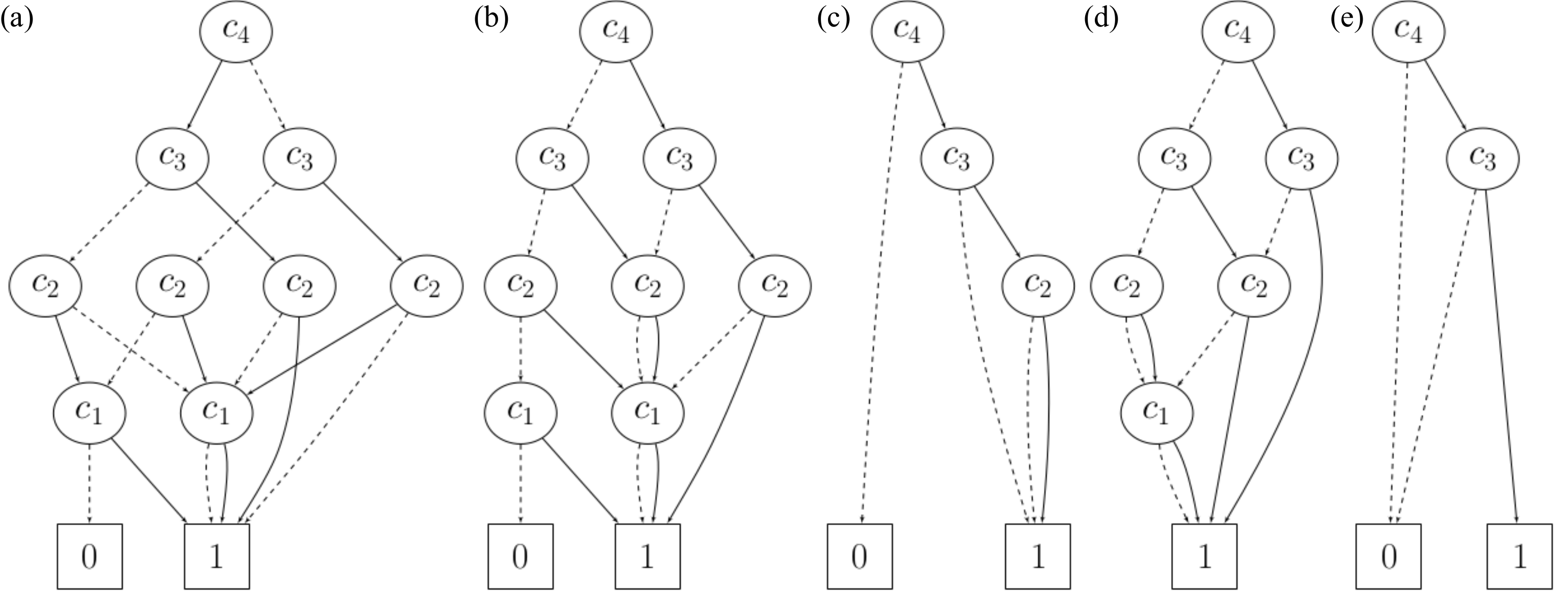}
    \caption{
      ZDDs of the two-dimensional example representing optional constraints.
      (a) ZDD representing the constraint for eliminating superperiodic structures.
      (b) ZDD representing the constraint for eliminating incomplete structures that contain both 0 and 1 labels.
      (c) Isomorphism-eliminated ZDD in which superperiodic and incomplete labelings are eliminated.
      (d) ZDD representing the constraint at $x=0.5$.
      (e) Isomorphism-eliminated ZDD at $x=0.5$ in which superperiodic and incomplete labelings are eliminated.
    }
    \label{fig:zdd_additional_constraints}
\end{figure*}

We should finally emphasize that superperiodic and incomplete structures can be optionally eliminated in a similar way to the construction of isomorphism-eliminated ZDDs.
For simplicity, we describe additional constraints for eliminating superperiodic and incomplete structures in binary systems ($k=2$).
A generalization of the constraints to multicomponent systems ($k \geq 3$) is achieved in a straightforward manner.

A superperiodic structure is invariant for a translational symmetry operation, not those of the sublattice $L_{\mathbf{M}}$ including the identity operation $e$.
Therefore, the ZDD representing the constraint for eliminating superperiodic structures is written as
\begin{align}
  \mathcal{C}_{\mathrm{sp}} = \left\{ \mathbf{c} \in \{ 0, 1 \}^{|D_{\mathbf{M}}|} \mid \mathbf{c} \neq \sigma(\mathbf{c}) \, (\forall \sigma \in T_{\mathbf{M}} \backslash \{ e \}) \right\},
\end{align}
where $T_{\mathbf{M}}$ denotes the permutation group mapped from the translational operations in the space group $\mathcal{H}_{\mathbf{M}}$.
Owing to the analogy with \eqnref{eq:polya-problem}, the ZDD representing $\mathcal{C}_{\mathrm{sp}}$ can be constructed using a procedure similar to the one shown in Appendix~\ref{sec:frontier}.
Figure~\ref{fig:zdd_additional_constraints}~(a) illustrates the ZDD representing $\mathcal{C}_{\mathrm{sp}}$ for the two-dimensional supercell shown in Sec.~\ref{sec:unique-example} as an example.

An incomplete structure corresponds to labeling that does not contain all labels, such as $\mathbf{c} = (0, 0, 0, 0)$.
The ZDD representing the constraint for eliminating incomplete structures is given as
\begin{align}
  \mathcal{C}_{\mathrm{incomplete}} = \left\{ \mathbf{c} \in \{ 0, 1 \}^{|D_{\mathbf{M}}|} \mid \mathbf{c} \supseteq \{ 0, 1 \} \right\},
\end{align}
which is constructed similarly to the one-of-$k$ ZDDs introduced in multicomponent systems.
The ZDD of the two-dimensional example representing the constraint for eliminating incomplete structures is illustrated in Fig.~\ref{fig:zdd_additional_constraints}(b).

Then, the set of nonequivalent labelings that does not contain superperiodic and incomplete ones is the intersection of the ZDDs, expressed as
\begin{align}
  \label{eq:zdd_sp_incomplete}
  \mathcal{C}_{ \mathbf{M}, 2 } \cap \mathcal{C}_{\mathrm{sp}} \cap \mathcal{C}_{\mathrm{incomplete}}.
\end{align}
Figure~\ref{fig:zdd_additional_constraints}~(c) illustrates the intersection of the ZDDs in the two-dimensional example.
The derivative structures shown in Figs.~\ref{fig:square_lattice}~(c), (e), and (f) correspond to 1-paths of the intersection ZDD.

Furthermore, the present ZDD-based method can enumerate nonequivalent labelings with a given composition \cite{HART2012101}.
The ZDD representing labelings with the composition of label 1, $x$, is written as
\begin{align}
  \mathcal{C}_{x} = \left\{ \mathbf{c} \in \{ 0, 1 \}^{ | D_{\mathbf{M}} | } \mid \sum_{i=1}^{ |D_{\mathbf{M}}| } c_{i} = |D_{\mathbf{M}}| x \right\}.
\end{align}
Figure.~\ref{fig:zdd_additional_constraints}~(d) shows the ZDD representing labelings at $x = 0.5$ in the two-dimensional example.
Finally, the set of nonequivalent labelings eliminated superperiodic and incomplete labelings is obtained by
\begin{align}
  \label{eq:zdd_sp_incomple_concentration}
  \mathcal{C}_{ \mathbf{M}, 2 } \cap \mathcal{C}_{\mathrm{sp}} \cap \mathcal{C}_{\mathrm{incomplete}} \cap \mathcal{C}_{x}.
\end{align}
Figure~\ref{fig:zdd_additional_constraints}~(e) illustrates the ZDD representing \eqnref{eq:zdd_sp_incomple_concentration} in the two-dimensional example at $x = 0.5$.
As found in Fig.~\ref{fig:zdd_additional_constraints}~(e), only the 1-path is $\mathbf{c} = (1, 1, 0, 0)$ in the two-dimensional example at $x = 0.5$.


%

\section{\label{sec:conclusion}Conclusion}
We have proposed an efficient procedure with a compact data structure of ZDD to enumerate derivative structures or nonequivalent labelings for given lattice and sites.
We have applied the ZDD-based procedure to the enumeration of binary, ternary, and quaternary derivative structures from the simple fcc and hcp structures.
The present ZDD-based method can be easily applied to the structure enumeration derived from the other structures, and significantly increases the possible number of derivative structures to be enumerated.
This is as many as approximately $10^{17}$, which is $10^{7}$ times larger than the possible number of structures in the previous method.
In addition to the derivative structure enumeration, ZDD and similar approaches should be powerful tools for solving combinatorial problems in physics and materials science that can be reformulated as the enumeration of subgraphs from a given graph.


%

\begin{acknowledgments}
This work was supported by
  a Grant-in-Aid for Scientific Research (B) (Grant Number 19H02419),
  a Grant-in-Aid for Challenging Research (Exploratory) (Grant Number 18K18942),
  and a Grant-in-Aid for Scientific Research on Innovative Areas (Grant Number 19H05787)
from the Japan Society for the Promotion of Science (JSPS).
TH acknowledges 
  a Grant-in-Aid for Scientific Research (C) (Grant Number 18K11153)
  and a Grant-in-Aid for Scientific Research (S) (Grant Number 15H05711)
from JSPS.
\end{acknowledgments}


\section{Data Availability}
The data that support the findings of this study are available from the corresponding author upon reasonable request.

\appendix

\section{\label{sec:appendix-polya}P\'olya's counting theorem}

P\'olya's counting theorem derives the number of $k$-ary nonequivalent labelings for the permutation group $\Sigma_{\mathbf{M}}$.
The number of nonequivalent labelings $\mathcal{C}_{\mathbf{M}, k}$ defined in \eqnref{eq:polya-problem}, $| \mathcal{C}_{\mathbf{M}, k}|$, is given by \cite{polya1937,Polya:1987:CEG:26181,tucker1984applied}
\begin{equation}
    \label{eq:polya}
    | \mathcal{C}_{\mathbf{M}, k}|
    = \frac{1}{|\Sigma_{\mathbf{M}}|} \sum_{ \sigma \in \Sigma_{\mathbf{M}} } k^{ t_{1}(\sigma) + t_{2}(\sigma) + \dots + t_{|D_{\mathbf{M}}|}(\sigma) },
\end{equation}
where $t_{j}(\sigma)$ denotes the number of cycles with length $j$ in permutation $\sigma$.
The sequence $\mathbf{t} (\sigma) = (t_{1}(\sigma), \dots, t_{D_{\mathbf{M}}}(\sigma) ) $ is the type of permutation $\sigma$ \cite{Polya:1987:CEG:26181}, and the sum of the elements in the type indicates the total number of cycles in permutation $\sigma$.
In the two-dimensional example shown in \secref{sec:unique-example}, the permutations are expressed using the cycle notation as
\begin{eqnarray*}
    \sigma_{1} &=&
    \begin{pmatrix}
        1 & 2 & 3 & 4 \\
        1 & 2 & 3 & 4
    \end{pmatrix}
    = (1)(2)(3)(4)
    \\
    \sigma_{2} &=&
    \begin{pmatrix}
        1 & 2 & 3 & 4 \\
        2 & 3 & 4 & 1
    \end{pmatrix}
    = (1234)
    \\
    \sigma_{3} &=&
    \begin{pmatrix}
        1 & 2 & 3 & 4 \\
        3 & 4 & 1 & 2
    \end{pmatrix}
    = (13)(24)
    \\
    \sigma_{4} &=&
    \begin{pmatrix}
        1 & 2 & 3 & 4 \\
        4 & 1 & 2 & 3
    \end{pmatrix}
    = (4321)
    \\
    \sigma_{5} &=&
    \begin{pmatrix}
        1 & 2 & 3 & 4 \\
        1 & 4 & 3 & 2
    \end{pmatrix}
    = (1)(3)(24)
    \\
    \sigma_{6} &=&
    \begin{pmatrix}
        1 & 2 & 3 & 4 \\
        3 & 2 & 1 & 4
    \end{pmatrix}
    = (2)(4)(13)
    \\
    \sigma_{7} &=&
    \begin{pmatrix}
        1 & 2 & 3 & 4 \\
        2 & 1 & 4 & 3
    \end{pmatrix}
    = (12)(34)
    \\
    \sigma_{8} &=&
    \begin{pmatrix}
        1 & 2 & 3 & 4 \\
        4 & 3 & 2 & 1
    \end{pmatrix}
    = (14)(23).
\end{eqnarray*}
Therefore, the types of permutation are derived as
\begin{eqnarray*}
    \mathbf{t}(\sigma_{1}) &=& (4, 0, 0, 0) \\
    \mathbf{t}(\sigma_{2}) &=& (0, 0, 0, 1) \\
    \mathbf{t}(\sigma_{3}) &=& (0, 2, 0, 0) \\
    \mathbf{t}(\sigma_{4}) &=& (0, 0, 0, 1) \\
    \mathbf{t}(\sigma_{5}) &=& (2, 1, 0, 0) \\
    \mathbf{t}(\sigma_{6}) &=& (2, 1, 0, 0) \\
    \mathbf{t}(\sigma_{7}) &=& (0, 2, 0, 0) \\
    \mathbf{t}(\sigma_{8}) &=& (0, 2, 0, 0),
\end{eqnarray*}
and the number of nonequivalent labelings is given by
\begin{equation}
    | \mathcal{C}_{\mathbf{M}, k}| = \frac{1}{8} \left( k^{4} + 2 k^{3} + 3 k^{2} + 2k \right).
\end{equation}


%
\section{\label{sec:frontier}Frontier-based method for deriving isomorphism-eliminated ZDD}

\begin{figure*}[tb]
    \centering
    \includegraphics[width=\linewidth]{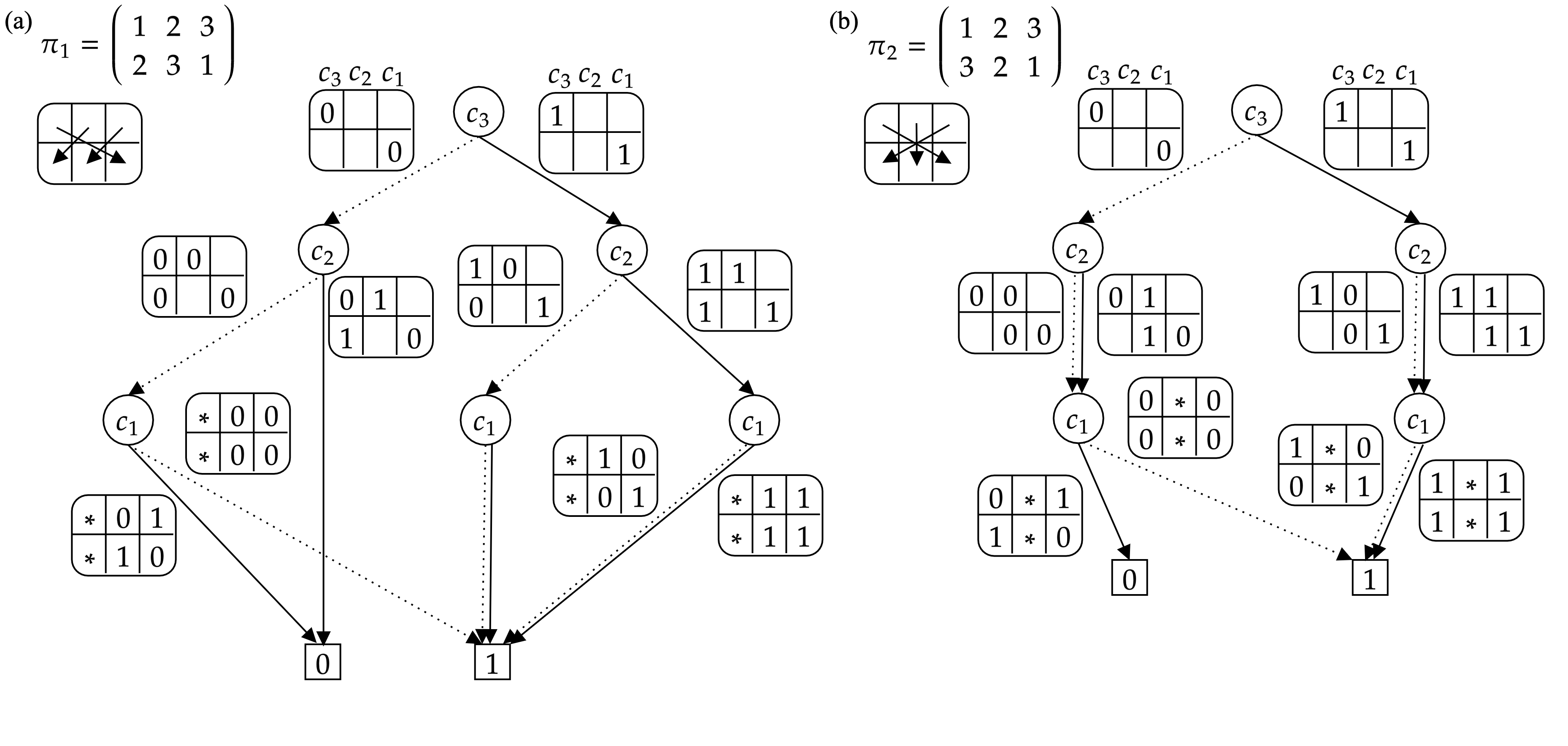}
    \caption{
        Development process of isomorphism-eliminated ZDDs with three labels, $\mathcal{C}_{\mathbf{M},2}^{(\pi)}$, for permutations $\pi_1$ and $\pi_2$.
        The solid and dotted arrows indicate 1-edges and 0-edges, respectively.
        The order of variables is fixed as $c_{3}$, $c_{2}$, and $c_{1}$.
        To show the development process of ZDDs, partially determined labeling and its permuted labeling are also shown for each edge.
        The upper and lower blocks indicate $(c_3, c_2, c_1)$ and $(c_{\pi(3)}, c_{\pi(2)}, c_{\pi(1)})$, respectively.
        The asterisks indicate discarded labels that are not needed in a later comparison process.
    }
    \label{fig:example-lexcographic-elimination}
\end{figure*}

Regarding the derivation of a ZDD, we lose the greatest advantage of ZDD once we have constructed the complete binary decision tree.
Therefore, primitive set operations between two ZDDs based on Bryant's algorithm have been used to derive ZDDs \cite{bryant1986graph,sasao2014applications}.
Recently, frontier-based methods improving the efficiency of deriving ZDDs have been proposed.
They are specially developed for various graph-based enumeration problems including the enumeration of $s$-$t$ paths \cite{knuth2009art,KAWAHARA2017} and spanning trees \cite{inoue2016graphillion}.

The frontier-based method is a dynamic programming method that uses specific structural properties of a given graph \cite{sasao2014applications}.
Therefore, the algorithm design of the frontier-based method strongly depends on the target problem.
Heuristic rules have been used in the algorithm; hence, the frontier-based method does not necessarily derive the irreducible ZDD.
Nonetheless, a well-designed algorithm of the frontier-based method is known to enable the building of a much more compressed data structure than the original binary decision tree.

The present frontier-based method was proposed to enumerate all non-isomorphic subgraphs of a given graph with respect to the automorphism of the graph by one of the authors of this study \cite{Horiyama2018}.
It was applied to the enumeration of all non-isomorphic developments of Platonic and Archimedean solids and $d$-dimensional hypercubes.
This method can be applied to the enumeration of binary derivative structures ($k=2$) in a straightforward manner because it can be regarded as such a subgraph enumeration problem.
Figure~\ref{fig:example-lexcographic-elimination} shows the development process of isomorphism-eliminated ZDDs with three labels, $\mathcal{C}_{\mathbf{M},2}^{(\pi)}$, for permutations
\begin{equation}
    \pi_{1} =
    \begin{pmatrix}
        1 & 2 & 3 \\
        2 & 3 & 1
    \end{pmatrix}, \:\:
    \pi_{2} =
    \begin{pmatrix}
        1 & 2 & 3 \\
        3 & 2 & 1
    \end{pmatrix}.
\end{equation}
The key ideas to efficiently construct an isomorphism-eliminated ZDD, $\mathcal{C}_{\mathbf{M},2}^{(\pi)}$, within a frontier-based method are
(1) comparing partially determined labeling and its permuted labeling and
(2) retaining only frontier labels that should be compared in a later process, not all the labels that are already determined.

The following branching and sharing rules are applied to derive ZDDs.
(1) If the relationship $\mathbf{c} \nsucceq \mathbf{\pi(c)}$ is decided in the comparison of partially determined labeling and its permuted labeling, all paths containing the partially determined labeling are never solutions.
The edge corresponding to the partially determined labeling is directly connected to the 0-terminal node.
An example of such a branching is $\mathbf{c} = (0\:1\:\circ)$ and $\mathbf{\pi(c)} = (1\:\circ\:0)$ for permutation $\pi_1$, where $\circ$ denotes the undetermined label.
(2) If the relationship $\mathbf{c} \succeq \mathbf{\pi(c)}$ is decided in the comparison of partially determined labeling and its permuted labeling, all paths containing the partially determined labeling are solutions.
The edge is connected to a ZDD that corresponds to the binary decision tree where every terminal node has the value of one.
An example of such a branching is $\mathbf{c} = (1\:0\:\circ)$ and $\mathbf{\pi(c)} = (0\:\circ\:1)$ for permutation $\pi_1$.
(3) If neither of the above relationships is decided, the labels used in the comparison are removed from the set of frontier labels that should continue to be compared.
They do not need to be compared in a later process after comparing label $c_{i}$ and permuted label $c_{\pi(i)}$ for site $i$.
Its example is $\mathbf{c} = (0\:0\:\circ)$ and $\mathbf{\pi(c)} = (0\:\circ\:0)$ for permutation $\pi_1$.
Labels $c_1$ and $c_{\pi(1)}$ are then discarded, and we denote them as $\mathbf{c} = (*\:0\:\circ)$ and $\mathbf{\pi(c)} = (*\:\circ\:0)$, where $*$ indicates a discarded label.
(4) Two nodes are merged when the values of their frontier labels coincide \footnote{
  Rigorously, two nodes are regarded as equivalent if their labels that are already compared coincide in addition to their frontier labels (see Ref.~\onlinecite{Horiyama2018}).
}.


%
\section{\label{sec:memory} Memory consumption}

\begin{figure*}[tb]
    \centering
    \includegraphics[width=\linewidth]{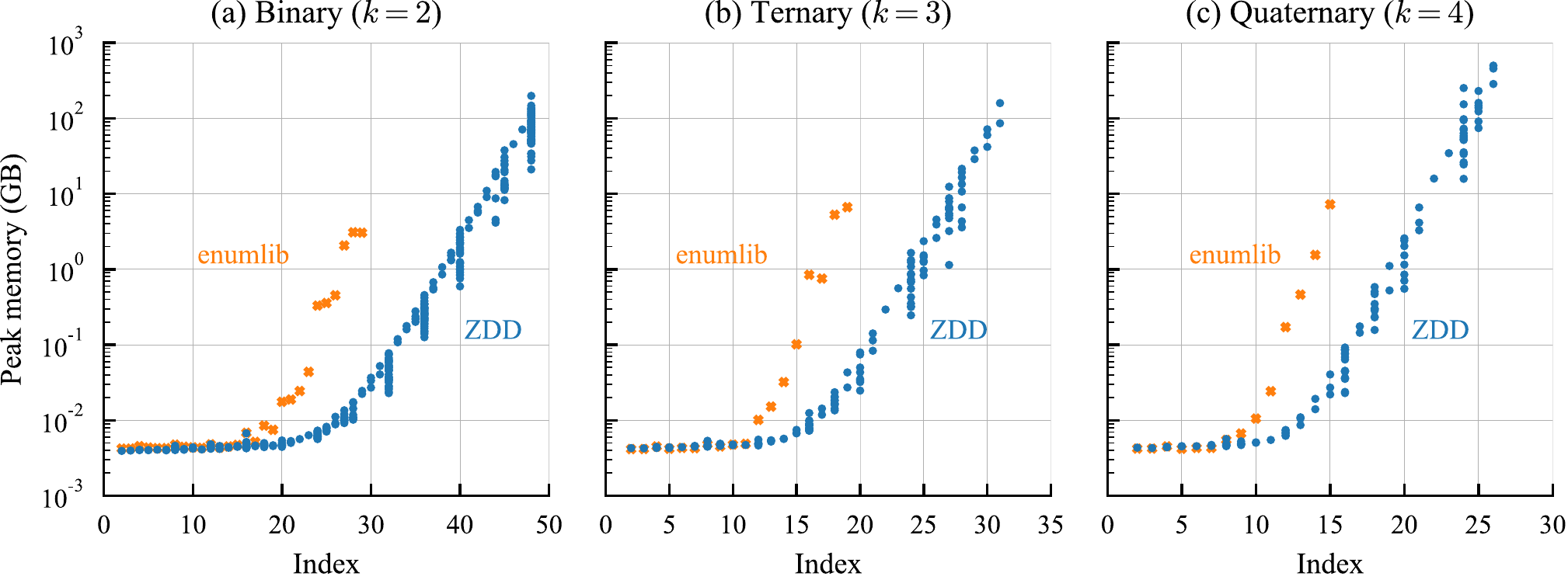}
    \caption{
        (Color online)
        Dependence of peak memory required to enumerate fcc derivative structures on the supercell index using the present ZDD-based method, shown by the blue closed circles.
        For comparison, the peak memory of the previous method to perform \software{enumlib} is also shown by the orange crosses.
        The left, middle, and right panels show the peak memory required to enumerate fcc derivative structures for binary, ternary, and quaternary systems, respectively.
    }
    \label{fig:each_zdd_vs_enumlib_fcc_memory}
\end{figure*}

Regarding memory consumption in the construction of ZDDs, we practically should consider not only the size of the allocated memory of the final ZDD representing a set of nonequivalent labelings but also the peak memory required for allocating a ZDD representing a set of lexicographically larger labelings for a permutation.
Figure~\ref{fig:each_zdd_vs_enumlib_fcc_memory} shows the memory required for the ZDD method in enumerating fcc nonequivalent labelings compared with that required for the previous method implemented in \software{enumlib} \cite{Hart2008,Hart2009,Morgan2017,enumlib}.
We show the peak memory required for constructing a ZDD for a non-isomorphic permutation group, while we show the peak memory required for enumerating derivative structures for a given index in the previous method.
As can be seen in Fig.~\ref{fig:each_zdd_vs_enumlib_fcc_memory}, the required memory for the ZDD method increases more slowly with the increase in the index than that for the previous method.
For example, the peak memory consumed at the index of 29 is approximately $3.0~\mathrm{GB}$ in the previous method and $25~\mathrm{MB}$ in the ZDD-based method.


%

\bibliography{references}%

\end{document}